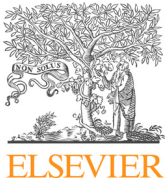
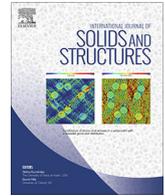

# Multi-scale modelling of concrete structures affected by alkali-silica reaction: Coupling the mesoscopic damage evolution and the macroscopic concrete deterioration

E.R. Gallyamov [a],[*], A.I. Cuba Ramos [a], M. Corrado [b], R. Rezakhani [c], J.-F. Molinari [a]

[a] Civil Engineering Institute, Materials Science and Engineering Institute, École Polytechnique Fédérale de Lausanne (EPFL), Station 18, CH-1015 Lausanne, Switzerland
[b] Department of Structural, Geotechnical and Building Engineering, Politecnico di Torino, Corso Duca degli Abruzzi 24, 10129 Torino, Italy
[c] Department of Mechanical Engineering and Material Science, Duke University, Durham, USA

**ABSTRACT**

A finite-element approach based on the first-order FE[2] homogenisation technique is formulated to analyse the alkali-silica reaction-induced damage in concrete structures, by linking the concrete degradation at the macro-scale to the reaction extent at the meso-scale. At the meso-scale level, concrete is considered as a heterogeneous material consisting of aggregates embedded in a mortar matrix. The mechanical effects of the Alkali-Silica Reaction (ASR) are modelled through the application of temperature-dependent eigenstrains in several localised spots inside the aggregates, and the mechanical degradation of concrete is modelled using continuous damage model, which is capable of reproducing the complex ASR crack networks. Then, the effective stiffness tensor and the effective stress tensor for each macroscopic finite element are computed by homogenising the mechanical response of the corresponding representative volume element (RVE). Convergence between macro- and meso-scales is achieved via an iterative procedure. A 2D model of an ASR laboratory specimen is analysed as a proof of concept. The model is able to account for the loading applied at the macro-scale and the ASR-product expansion at the meso-scale. The results demonstrate that the macroscopic stress state influences the orientation of damage inside the underlying RVEs. The effective stiffness becomes anisotropic in cases where damage is aligned inside the RVE.

## 1. Introduction

The alkali-silica reaction (ASR) is the most common type of alkali-aggregate reaction, which is the generic term for reactions between the alkaline concrete pore solution and certain mineral phases within the aggregates. At the scale of the aggregates and cement paste, ASR manifests itself in the form of local silica dissolution, growth of micro-cracks, their filling with the ASR products, and the overall expansion of aggregates and eventually paste. Micro-cracks are homogeneously distributed within the meso-structure. Their orientation shows dependence on the stress state (Dunant et al., 2012); typical crack patterns vary from randomly oriented to echelons of co-directional cracks. Damage development results in a loss of stiffness and tensile strength (see Fig. 1). The nature of the ASR expansion is subject of debating. Significant alkali-aggregate expansions are only observed, when the relative humidity exceeds 80% (Stark, 1991). The most popular hypothesis regarding the expansion of concrete is swelling of the ASR product after water absorption (Swamy, 1992). The problem is further complicated by the presence of two types of ASR product: amorphous and crystalline. Shi et al. (2019) recently showed that the synthesised crystalline ASR product has relatively low water uptake capacity, thereby reducing swelling to negligible amounts.

In order to study the physical correlation between the gel formation and cracking, several meso- and micro-scale models have been developed (Bazant et al., 2000, Multon et al., 2009, Comby-Peyrot et al., 2009, Alnaggar et al., 2013, Esposito, 2016, etc.). These models have proven to be useful in the analysis of laboratory experiments of ASR, because a systematic variation of single input parameters is possible.

Results of the laboratory experiments on ASR expansion of concrete under loading (Larive, 1998; Multon and Toutlemonde, 2006; Gautam et al., 2016) show the so-called "stress-induced anisotropy" - restriction of the chemical expansion in the direction of compressive load and its partial transfer to the perpendicular direction. Highly compressed specimens show negligible expansion in the direction of load. The anisotropy was partially captured by Dunant and co-workers (Dunant et al., 2010; Dunant et al., 2012), who developed a two-dimensional meso-scale model with explicit representation of ASR-gel pockets and evolution of damage at the meso-scale. However, the model failed at high loads by showing specimens contraction along the load direction, which is never the case in the experiments. Later, this model was complemented by Giorla et al. (2015) by adding visco-elasticity to the cement paste and studying its stress relaxation effect. Schlangen et al. (2010) used a similar concept of *one-by-one* damaging events to simulate ASR degradation of concrete with a 3D lattice model.

* Corresponding author.
*E-mail address:* emil.gallyamov@epfl.ch (E.R. Gallyamov).

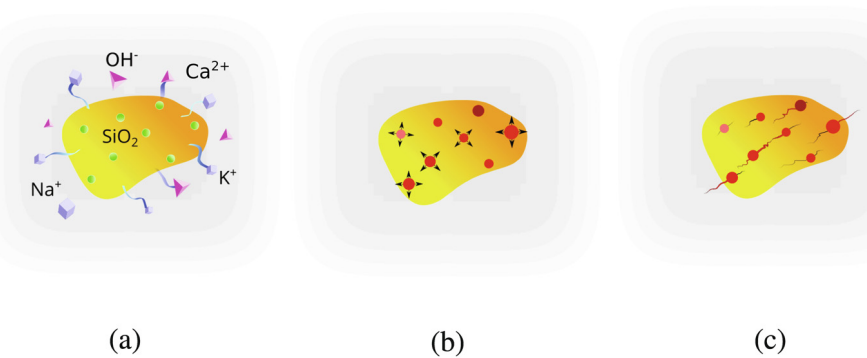

**Fig. 1.** Stages of concrete damaging due to ASR: (a) alkalis from cement paste diffuse into aggregates and react with amorphous silica; (b) resulting ASR products exert internal pressure on aggregates; (c) tensile cracking advance into aggregates and cement paste.

Since ASR is most commonly detected in large concrete structures exposed to water, such as sea walls, bridge piers and dams, several macro-models that can address the consequences of ASR directly at the structural scale of concrete have also been proposed. These models aim at predicting when and how an affected structure should be repaired. Such predictions are important in order to avoid a significant reduction in service life of the structure and hence help alleviate the economical effects of ASR. In macro-scale models, concrete is considered as a homogeneous material and the underlying meso-structure consisting of aggregates and the cement paste is not resolved. This simplified material description is necessary because an explicit representation of the aggregates and the cement paste at this scale would yield to computational models of unacceptable size.

One large subgroup of these macro-scale models is based on a phenomenological approach (Larive, 1998; Capra and Bournazel, 1998; Ulm et al., 2000; Saouma et al., 2006; Martin et al., 2012; Omikrine et al., 2014, etc.). These models compute directly the macroscopic ASR strain by extrapolating results from laboratory experimental measurements to field conditions. Phenomenological models require extensive experimental studies to determine all the input parameters needed for the simulations.

Another group of macro-scale ASR models incorporates poromechanics, in which concrete is considered as a solid skeleton permeated by open porosity (Capra and Sellier, 2003; Bangert et al., 2004; Comi et al., 2009; Grimal et al., 2010; Pignatelli et al., 2013; Multon and Sellier, 2016; Esposito, 2016, etc.). Constitutive laws for macro-scale simulations are derived from homogenisation principles. The ASR-induced damage is typically assumed to be isotropic. The solid phase does not distinguish between aggregates and cement paste, which causes a uniform damage distribution inside the concrete skeleton.

Recently, due to the advancement of high-performance computing, the first numerical multi-scale models have been proposed for ASR. These models couple numerical computations at the large and fine scales in a single simulation. The multi-scale approach is advantageous because the consequences of ASR at a larger scale can be computed directly from the reaction advancement inside an underlying representative volume element (RVE). The RVE represents the heterogeneous structure of concrete at a smaller length scale. A model by Puatatsananon et al. (2013) accounts for the diffusion processes and ASR-gel swelling at the level of aggregates. Wu et al. (2014) proposed a model that couples gel production and expansion in micropores of hardened cement paste (HCP) to its manifestation at the meso-scale. However, both models are lacking explicit representation of ASR-induced cracks which could help to reproduce expansion anisotropy. A 3D ASR-model by Rezakhani et al. (2019) couples the lattice discrete-particle model (LDPM) employed at the meso-scale to the macro-scale problem resolved by the finite element method (FEM). In their approach, authors approximate swelling of single ASR-product pockets by overall expansion of aggregates. This advanced model captures anisotropic ASR expansion under loading and produces localised crack patterns. On the other hand, cracks are only allowed to develop on plane shapes delimited by the aggregates centres, which is not the case of a real ASR-affected concrete.

In order to study the consequences of ASR in large structures such as concrete dams, multi-scale models that establish a link between the meso- and the macro-scale are required. The development of such a multi-scale model for the ASR-affected concrete, which would naturally capture the expansion anisotropy effect, is the objective of the present work. In the course of this study, special attention is paid to the improvement of the meso-mechanical model of ASR by adding physics-based concepts. Those include material strain-softening, cracks opening and closure, and explicit representation of the ASR-product sites. Nonetheless, some aspects of the model remain phenomenological, such as the ASR-product expansion law.

The future usage of such model is meant for dam engineers and owners in order to understand the interplay between the mesoscopic state of concrete and macroscopic behaviour of a structure. The level of details at the meso-scale facilitates visualisation of phenomena at hand, which is hindered by pure macroscopic models. Apart from explaining the current behaviour of an ASR-affected structure, given enough field measurements for calibration, the model could be used for predictions of the structure's future behaviour. Such opportunity is crucial for the dam owning companies as it allows to make well-timed planning and preview necessary interventions.

The remainder of the paper is organised as follows: in Section 2, the ASR multi-scale model is described in detail, including parallel implementation. Section 3 introduces the constitutive law used at the meso-scale. Calibration of the proposed meso-scale model is described in Section 4. Subsequently, in Section 5, the multi-scale model is applied to a 2D cross-section of laboratory concrete specimens in order to verify the method and to highlight the capabilities of the proposed approach. We discuss how the orientation of damage inside the RVE and the anisotropy of the effective stiffness at the macro-scale are a natural consequence of the interaction between the meso- and the macro-scale.

## 2. The ASR multi-scale model

Our objective is the modelling of ASR at the macro-scale by incorporating the results of meso-scale simulations. In this work, we use a first-order FE$^2$ homogenisation scheme to couple the macro- and the meso-scale. This method consists of two coupled finite element (FE) problems, one for the macro-scale and the other for the fine scale, *i.e.* the scale of the underlying RVE, which determines the effective material behaviour. Note that in this work, the fine scale of the FE$^2$ problem is the meso-scale of concrete. The FE$^2$ method is based on the principle of *separation of scales*, which has been formulated as *"The microscopic length scale is assumed to be much smaller than the characteristic length over which the macroscopic loading varies in space"* by Geers et al. (2010). This assumption is valid for the multi-scale modelling of ASR, since large concrete structures have several tens of meters of width and height, and the meso-scale RVE is in the centimetre range. The concept of a first-order FE$^2$ homogenisation scheme for small deformations is illustrated in Fig. 2.

Note that in the following, the subscripts $M$ and $m$ will be used to denote physical quantities at the macro- and the meso-scale, respectively. For instance, $\boldsymbol{\sigma}_M$ is the macroscopic stress for which the mesoscopic counterpart is $\boldsymbol{\sigma}_m$. Every Gauss integration point of the macro-scale FE problem is coupled to a meso-structural RVE of concrete. Quasi-static conditions are assumed at both scales because the advancement of ASR is slow. The two scales are coupled through the macroscopic deformation gradient $\boldsymbol{F}_M$, the effective stiffness tensor $\mathbb{C}_M$, and the macroscopic stress $\boldsymbol{\sigma}_M$. The boundary conditions of the meso-scale boundary value problem (BVP) are a function of $\boldsymbol{F}_M$. The BVP is solved for the given boundary conditions, and the effective stiffness tensor $\mathbb{C}_M$ and the homogenised mesoscopic stress $\boldsymbol{\sigma}_M$ are then computed and passed back to the macro-scale. The balance between the internal and external forces at the macro-scale is verified. In the case of imbalance, the whole procedure is repeated in the next iteration. As a convergence criterion, we use the norm of the difference between the internal and external forces, also known as *residual*.

In the course of ASR development within RVEs, cracks may coalesce and therefore violate the separation of scales' principle. The strain localisation limits the concept of homogenisation (Coenen et al., 2012). We test the validity of the numerical homogenisation for a problem where cracks are evenly distributed over the meso-scale volume. The following model components are called to extend the applicability of the classical homogenisation concept to the ASR case:

- An extended constitutive model of concrete. The material model adopted in this study is the continuous orthotropic damage with stiffness recovery upon crack closure, which will be described in Section 3.
- A robust stiffness homogenisation procedure. This measure is called to overcome the ill-posedness of the macro-scale boundary value problem. This is done by an adaptive homogenisation of stiffness either by tension or by compression tests, which produce a non-singular stiffness tensor. The resulting tensors are better suited for the macroscopic stress state, which also improves the convergence rate of the iterative scheme. The homogenisation procedure is described in Subsection 2.4.

One distinctive feature of the current multi-scale model is the fact that the external loading is not only coming from the macro-scale, but also from the fine scale in the form of expanding ASR product. The interplay between these two load scales is interesting both from the physical and the numerical point of view.

The proposed multi-scale model is two-dimensional. This limitation is caused by its high computational cost, which mainly comes from the fine resolution of RVEs and the adopted way of solving the BVP. However, the concept of the proposed approach is universal, and the model could be easily extended to 3D given enough computational powers. Similar to many other 2D multi-scale models (Erinc et al., 2013; Iskhakov et al., February 2018; Puatatsananon et al., 2013, etc.), we aim to study the physics at hand while being aware of the possible limitations.

### 2.1. Macro-scale problem

Let $\Omega_M$ denote a body of ASR-affected concrete at the macro-scale. The domain is bounded by $\Gamma_M$, which can be decomposed into the Neumann boundary $\Gamma_{M,t}$ and the Dirichlet boundary $\Gamma_{M,u}$, such that the following definitions hold:

$$\Gamma_M = \Gamma_{M,u} \cup \Gamma_{M,t}, \tag{1}$$
$$\Gamma_{M,u} \cap \Gamma_{M,t} = \varnothing. \tag{2}$$

The principle of virtual work reads as:

$$\delta W_M = \int_{\Omega_M} \boldsymbol{\sigma}_M : \delta \boldsymbol{\varepsilon}_M \, d\Omega - \int_{\Omega_M} \boldsymbol{b}_M \cdot \delta \boldsymbol{u}_M \, d\Omega - \int_{\Gamma_{M,t}} \boldsymbol{t}_M \cdot \delta \boldsymbol{u}_M \, d\Gamma = 0, \tag{3}$$

where $\delta W_M$, $\boldsymbol{b}_M$ and $\boldsymbol{t}_M$ denote the macroscopic virtual work, macroscopic body force and macroscopic traction, and $\delta \boldsymbol{\varepsilon}_M$ and $\delta \boldsymbol{u}_M$ the macroscopic virtual strain and macroscopic virtual displacement. No assumptions are introduced regarding $\boldsymbol{\sigma}_M$. Instead, it is obtained

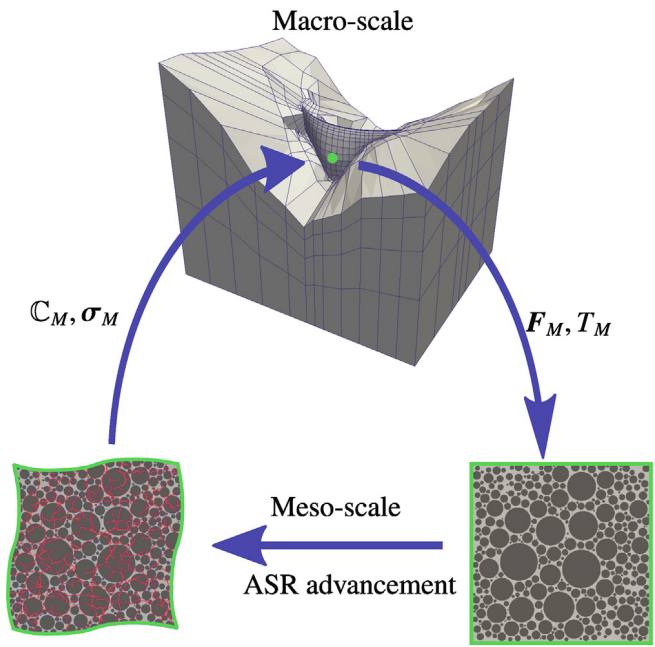

**Fig. 2.** Schematic illustration of the numerical homogenisation scheme for ASR simulations. Every macroscopic computational point is coupled to an underlying RVE. The boundary conditions for the RVE are defined through the macroscopic deformation gradient $F_M$. Macroscopic temperature $T_M$ is passed to the meso-scale for predicting the ASR product expansion rate. After solving the RVE problem, the macroscopic stress $\boldsymbol{\sigma}_M$ and the effective stiffness tensor $\mathbb{C}_M$ are computed.

directly from the meso-scale computations as explained in Subsections 2.4 and 2.5.

*2.2. Macro-to-meso transition*

The displacement field inside the RVE can be decomposed into two parts:

$$\boldsymbol{u}_m(\boldsymbol{x}) = \boldsymbol{u}_M + \hat{\boldsymbol{u}}(\boldsymbol{x}), \tag{4}$$

where $\hat{\boldsymbol{u}}(\boldsymbol{x})$ are micro-fluctuations. The current macroscopic state enters the RVE computations via the boundary conditions. In this work periodic boundary conditions are chosen because they give better estimates for the effective stiffness in comparison with the uniform displacement or uniform traction boundary conditions (Coenen et al., 2012). The periodic boundary conditions are defined as follows:

$$\boldsymbol{u}_{m,i} = (\boldsymbol{F}_M - \boldsymbol{1})\boldsymbol{x}_{m,i} \quad \text{for} \quad i = 1, 2, 3, 4 \tag{5a}$$

$$\boldsymbol{u}_{\Gamma_{m,34}} = \boldsymbol{u}_{\Gamma_{m,12}} + \boldsymbol{u}_{m,4} - \boldsymbol{u}_{m,1} \tag{5b}$$

$$\boldsymbol{u}_{\Gamma_{m,23}} = \boldsymbol{u}_{\Gamma_{m,14}} + \boldsymbol{u}_{m,2} - \boldsymbol{u}_{m,1} \tag{5c}$$

where $\boldsymbol{F}_M$ denotes the deformation gradient at the corresponding macroscopic material point, $\boldsymbol{u}_{m,i}$ is the displacement of the corner node $i$, and $\Gamma_{m,12}, \Gamma_{m,23}, \Gamma_{m,34}$ and $\Gamma_{m,14}$ are the boundaries of the RVE (see Fig. 3). While the terms $[\boldsymbol{u}_{m,4} - \boldsymbol{u}_{m,1}]$ and $[\boldsymbol{u}_{m,2} - \boldsymbol{u}_{m,1}]$ represent the macroscopic component of displacement, $\boldsymbol{u}_{\Gamma_{m,12}}$ and $\boldsymbol{u}_{\Gamma_{m,14}}$ are the periodic micro-fluctuations at the boundary pairs ($\Gamma_{m,12}$ and $\Gamma_{m,34}$) and ($\Gamma_{m,14}$ and $\Gamma_{m,23}$), correspondingly.

*2.3. ASR meso-scale problem*

The ASR meso-scale model presented here is purely mechanical. The dissolution and precipitation processes, which govern the evolution of the ASR-product, are not modelled. The effect of temperature is taken into account through the phenomenological ASR-product expansion law, which is discussed further. In the following, we summarise the basic equations and assumptions constituting the ASR meso-scale model.

We now consider the two-dimensional domain $\Omega_m$ with closure $\Gamma_m$, which represents an ASR-affected RVE as shown in Fig. 3. Three mutually exclusive phases constitute $\Omega_m$: the mortar $\Omega_{m,C}$, which includes the cement paste and the sand, the aggregates $\Omega_{m,A}$, and the ASR sites $\Omega_{m,G}$. An isotropic eigenstrain field $\boldsymbol{\varepsilon}_{m,eig}$ is imposed at each ASR site to account for the expansion. The eigenstrain is linked to the elastic strain $\boldsymbol{\varepsilon}_{m,el}$ via the following equation:

$$\boldsymbol{\varepsilon}_m = \boldsymbol{\varepsilon}_{m,el} + \boldsymbol{\varepsilon}_{m,eig}, \tag{6}$$

where $\boldsymbol{\varepsilon}_m$ is the infinitesimal strain tensor $\boldsymbol{\varepsilon}_m = \nabla_S \boldsymbol{u}_m = \frac{1}{2}(\nabla \boldsymbol{u}_m + \nabla \boldsymbol{u}_m^T)$. Note that the Cauchy stress $\boldsymbol{\sigma}_m$ depends only on the elastic part of the strain, i.e. $\boldsymbol{\sigma}_m = \boldsymbol{\sigma}_m(\boldsymbol{\varepsilon}_{m,el})$. The volume increase of the ASR product is approximated by increasing the imposed eigenstrain in every step of the simulation. In this study, we assume that the strain imposed to an ASR site $\boldsymbol{\varepsilon}_{m,eig}$ is proportional to the total amount of the generated ASR product which in turn is proportional to the chemical reaction extent $\xi$. Dependence of alkali-silica reaction kinetics on the temperature and the relative humidity was experimentally studied by Larive (1998). A first-order kinetic law in isothermal conditions results in the explicit equation for the chemical reaction extent $\xi$:

$$\xi(t, T) = \frac{1 - \exp[-t/\tau_{ch}(T)]}{1 + \exp[-t/\tau_{ch}(T) + \tau_{lat}(T)/\tau_{ch}(T)]}, \tag{7}$$

$$\tau_{ch}(T) = \tau_{ch}(T_0) \exp[U_C(1/T - 1/T_0)], \tag{8}$$

$$\tau_{lat}(T) = \tau_{lat}(T_0) \exp[U_L(1/T - 1/T_0)], \tag{9}$$

where $t$ and $T$ are the current time and temperature, $\tau_{lat}$ and $\tau_{ch}$ are latency and characteristic times either at the current, $T$, or reference, $T_0$, temperature, $U_C$ and $U_L$ are the activation energy constants. Therefore, the phenomenological ASR expansion law reads:

$$\boldsymbol{\varepsilon}_{eig}(t, T) = \boldsymbol{\varepsilon}(\infty) \frac{1 - \exp[-t/\tau_{ch}(T)]}{1 + \exp[-t/\tau_{ch}(T) + \tau_{lat}(T)/\tau_{ch}(T)]} \mathbb{I}, \tag{10}$$

where $\mathbb{I}$ is the identity matrix and $\boldsymbol{\varepsilon}(\infty)$ is the asymptotic volumetric expansion strain in the stress-free experiment (Larive, 1998; Ulm et al., 2000). The latency time $\tau_{lat}$, the characteristic time $\tau_{ch}$, thermal activation constants $U_C$ and $U_L$, and the asymptotic strain $\boldsymbol{\varepsilon}(\infty)$ are the calibration parameters of the model. Given a homogeneous temperature of a specimen, the proposed law results in the same expansion values at all ASR sites.

Quasi-static conditions are assumed because the ASR degradation process is slow and hence the expression for the virtual work is obtained:

$$\delta W_m = \int_{\Omega_m} \mathbb{C}_m (\boldsymbol{\varepsilon}_m - \boldsymbol{\varepsilon}_{m,eig})$$
$$: \delta \boldsymbol{\varepsilon}_m \, d\Omega - \int_{\Omega_m} b_m \delta u_m \, d\Omega - \int_{\Gamma_{m,t}} t_m \delta u_m \, d\Gamma = 0, \tag{11}$$

In this ASR meso-scale model the aggregates and the mortar are assumed to be quasi-brittle materials and the ASR product is modelled as linear elastic. Its expansion causes high stresses in the surrounding material. Compressive and tensile stresses are acting in radial and tangential direction respectively, as illustrated in Fig. 4. Because the tensile strength of the aggregates and the mortar is significantly lower than their compressive strength, it is valid to assume that the material fails locally always under tension with subsequent linear softening. The failure criterion and the constitutive law are depicted in Fig. 5. Material behaviour at the meso-scale is detailed in Section 3.

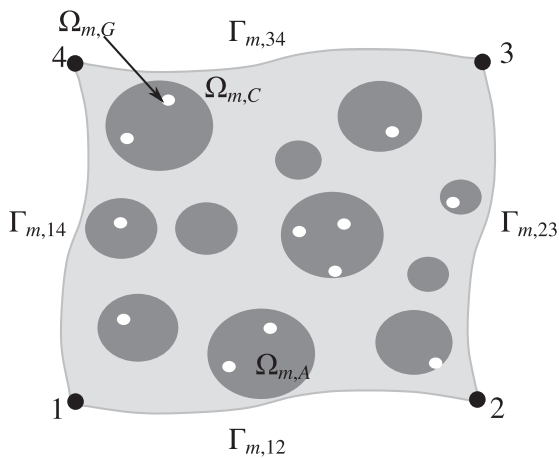

**Fig. 3.** Periodically deformed RVE of concrete with three mutually exclusive phases: mortar $\Omega_{m,C}$, aggregates $\Omega_{m,A}$ and ASR sites $\Omega_{m,G}$. The four edges are denoted by $\Gamma_{m,12}, \Gamma_{m,23}, \Gamma_{m,34}$ and $\Gamma_{m,14}$.

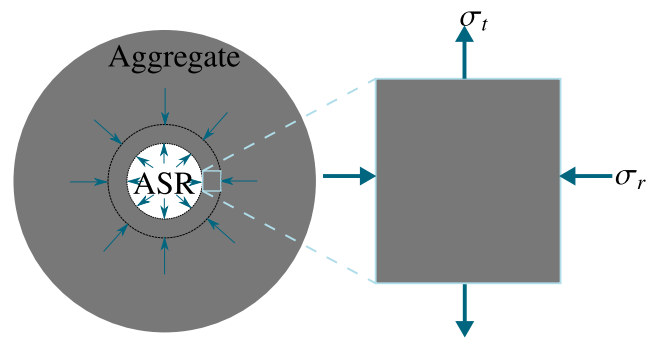

**Fig. 4.** Schematic representation of the stress state in the material around an ASR product site. Due to the volume increase of the ASR product the surrounding material experiences compressive stresses $\sigma_r$ in radial direction and tensile stresses $\sigma_t$ in tangential direction.

The computation of failure in brittle and quasi-brittle materials by means of the classical non-linear finite element method is often prone to numerical instabilities. In particular, for problems characterised by the simultaneous propagation of multiple cracks, as in the case of ASR, the solution of the system is likely to diverge and hence a completely different approach becomes necessary. Cuba Ramos et al. (2018) have shown that the use of the sequential linear analysis (SLA) (Rots, 2001; Rots et al., 2004; DeJong et al., 2008; Rots et al., 2008), allows to capture the complex crack networks inside the aggregates and the mortar. The idea of this method is to choose the load increments in such a way that for each imposed load increment, exactly one integration point undergoes softening and all other integration points remain below their failure criterion. The development of this method has been inspired by lattice models, in which the continuum is replaced by a lattice of beams. In lattice models, divergence problems do not occur because the solution is obtained from a sequence of linear analyses. In every step of the analysis, the beam with the highest load is detected and subsequently removed. The SLA transfers this approach to the continuum modelling. Consequently, in every step, the integration point with the highest load is detected, and if the damage criterion is satisfied, the stiffness and strength at this integration point are reduced according to the current damage value $d_i$, which is defined as

$$d_i = 1 - \frac{1}{a^i}, \qquad (12)$$

where the empirical reduction constant $a$, is brought to the power $i$, which is the number of the reduction step. In the following simulations, $a = 2$ and $i = 10$. After reaching this limit, the damage is brought to its maximum value (0.9999 in this case) and its update is stopped. Having a non-unity value of the maximum damage is necessary to avoid the stiffness matrix singularity.

In a recent work, Dunant et al. (2015) showed that the load scaling is not necessary for convergence when using the fracture criterion. It led to development of a modified version of SLA, where load was applied in a single step and multiple elements were allowed to undergo softening (Dunant et al., 2010). This modified SLA is adopted in our study.

After each damaging event, the mechanical properties referred to the damaged integration point are updated as follows:

$$E_i = E(1 - d_i), \quad v_i = v(1 - d_i), \quad \mu_i = \mu(1 - d_i), \qquad (13)$$

where the index denotes the reduction step. The tensile strength is reduced as

$$f_{t,i} = \varepsilon_u E_i \frac{E_t}{E_i + E_t} \qquad (14)$$

with

$$E_t = \frac{f_t}{\varepsilon_u - (f_t/E)}. \qquad (15)$$

Due to the discrete reduction of stiffness and strength, the stress-strain curve obtained with the SLA differs from the original softening curve, as illustrated in Fig. 6a. In order to ensure that the dissipated energy equals the theoretical value, different regularisation techniques can be applied (Rots et al., 2004). In this work, the combined regularisation technique, where the tensile strength $f_t$ and the ultimate strain $\varepsilon_u$ are simultaneously adjusted, is chosen and the resulting stress-strain curve is shown in Fig. 6b. Compared to lattice models, the advantage of the SLA is that the concept of strength, elasticity and fracture energy remain meaningful at the macro-scale. Because the use of the tangent stiffness is avoided in the SLA, numerical instabilities do not arise and, during the analysis, the equilibrium path of the material degradation is always followed. Another advantage is that due to the controlled increase of damage, jumps over the structural response do not occur. However, the major drawback of this method is its high computational cost. This cost results from the fact that a new solution has to be computed after each damage event. Consequently, the total number of computations of solutions is significantly larger than in a non-linear finite element approach. A more detailed description of the application of SLA to the meso-scale ASR problem can be found in the study by Cuba Ramos (2017).

### 2.4. Meso-to-macro transition

The stress distribution inside the RVE is obtained through the solution of the meso-scale boundary value problem. For the macro-scale, the average stress response of the RVE needs to be determined. This can be done using the *Hill-Mandel* macro-homogeneity condition (Hill, 1963):

$$\frac{1}{\Omega_m} \int_{\Omega_m} \boldsymbol{\sigma}_m : \boldsymbol{\varepsilon}_m \, d\Omega_m = \boldsymbol{\sigma}_M : \boldsymbol{\varepsilon}_M. \qquad (16)$$

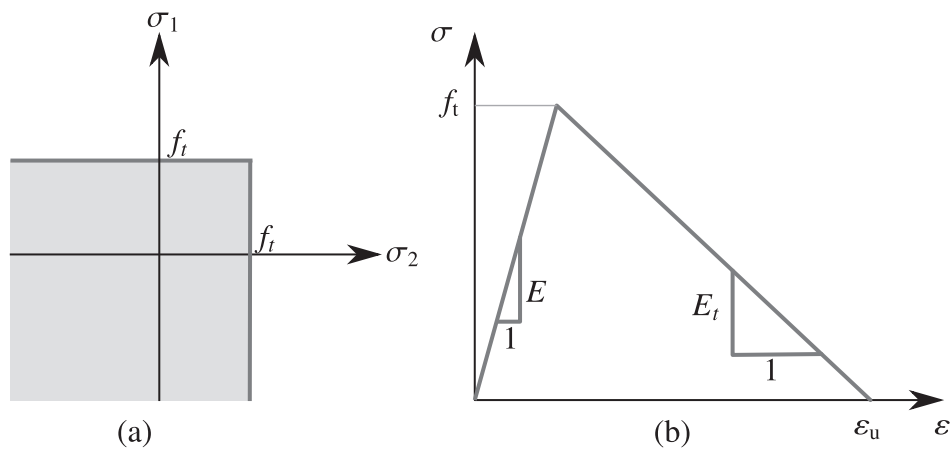

**Fig. 5.** Failure criterion and constitutive law for aggregates and mortar: (a) maximum principal tensile stress criterion. The material fails if the maximum principle tensile stress reaches the tensile strength limit $f_t$. (b) Bilinear law with an initial elastic loading phase and subsequent linear strain-softening.

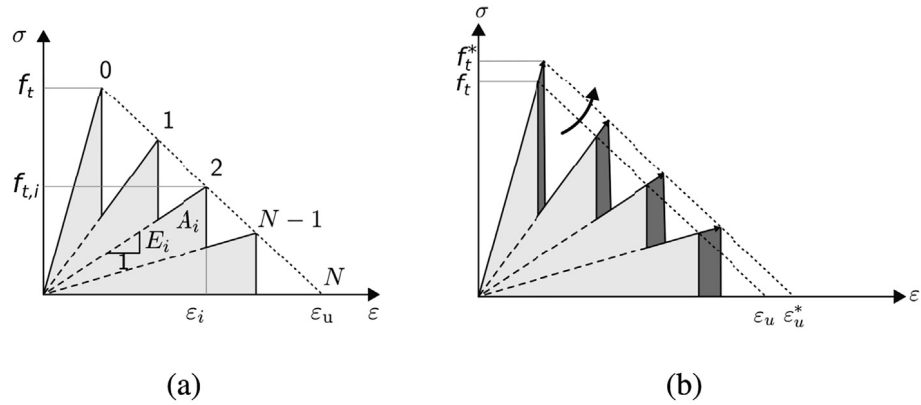

**Fig. 6.** Stress-strain curves obtained with the SLA: (a) curve without regularisation. The dissipated energy is less than the theoretical value. (b) Regularised stress-strain curve. The ultimate strain and the tensile strength are adjusted such that the dissipated energy equals the theoretical value.

Note that $\boldsymbol{\sigma}_m$ and $\boldsymbol{\varepsilon}_m$ are coupled to their macroscopic counterparts via the following equations:

$$\boldsymbol{\sigma}_m = \boldsymbol{\sigma}_M + \hat{\boldsymbol{\sigma}}, \tag{17}$$

$$\boldsymbol{\varepsilon}_m = \boldsymbol{\varepsilon}_M + \hat{\boldsymbol{\varepsilon}}. \tag{18}$$

where $\hat{\boldsymbol{\sigma}}$ and $\hat{\boldsymbol{\varepsilon}}$ are the micro-fluctuations of the stress and the strain fields, respectively. The *Hill-Mandel* macro-homogeneity condition implies that the work of the stress fluctuations $\hat{\boldsymbol{\sigma}}$ on the strain fluctuations $\hat{\boldsymbol{\varepsilon}}$ vanishes, i.e.

$$\frac{1}{\Omega_m} \int_{\Omega_m} \hat{\boldsymbol{\sigma}} : \hat{\boldsymbol{\varepsilon}} \, d\Omega_m = 0. \tag{19}$$

Therefore, $\hat{\boldsymbol{\sigma}}$ and $\hat{\boldsymbol{\varepsilon}}$ must be orthogonal to each other. The macroscopic stress can be expressed in terms of traction along the boundaries of the RVE:

$$\boldsymbol{\sigma}_M = \int_{\Gamma_m} t_m \otimes x_m \, d\Gamma_m. \tag{20}$$

It was shown by Suquet (1985) that the boundary conditions presented in Subsection 2.2 fulfil the *Hill-Mandel* condition and yield equivalence between the macroscopic stress and volume average of the mesoscopic stresses inside the RVE:

$$\boldsymbol{\sigma}_M = \int_{\Gamma_m} t_m \otimes x_m \, d\Gamma_m = \frac{1}{\Omega_m} \int_{\Omega_m} \boldsymbol{\sigma}_m \, d\Omega_m. \tag{21}$$

The above equation is used for the stress homogenisation at the meso-scale.

For the macroscale analysis, a tangent stiffness tensor is required at each integration point. The effective stiffness tensor is defined by the effective stress-strain relation

$$\boldsymbol{\sigma}_M = \mathbb{C}_M \boldsymbol{\varepsilon}_M. \tag{22}$$

Both, $\boldsymbol{\sigma}_M$ and $\boldsymbol{\varepsilon}_M$ are symmetric second-order tensors with three independent components in 2D. There are multiple methods of stiffness homogenisation available in the literature. The one proposed by Kouznetsova et al. (2001) comprises rearrangement of the full RVE stiffness tensor and its condensation with the help of position vectors of the fixed corner nodes. Miehe (1996) has suggested a method based on a forward difference approximation. In the present work, we adopt the virtual tests approach, the mathematical formulation of which is thoroughly described by Fritzen (2011). In the scope of this method, the results of three virtual loading tests, linearly independent from each other, are required to determine the symmetric fourth-order tensor $\mathbb{C}_M$.

For a specimen with isotropic stiffness reduction in cracks and no contact upon their closure, tensile and compressive loading tests would result in the same tensor $\mathbb{C}_M$. However, a material with orthotropic damage and stiffness recovery will behave differently under tension than under compression. While pulling a specimen would only open the cracks perpendicular to the load, its compression will close them and cause stresses in the bulk. This is however a simplification of a problem. In reality, cracks could be oriented in different directions, and the opening of one crack could be accompanied by the closure of another. That is why even the tensile test of a material with contact is expected to produce a higher stiffness than the similar test of a contactless material. In general, stiffness obtained via compression is higher than the one computed by tension. The effect of the test type on the stiffness reduction is demonstrated in Section 4.

Since the homogenised stiffness is used to solve the macro-scale problem for displacements, the better it approximates the stiffness of the meso-scale structure, the faster the multi-scale scheme converges. Providing low tensile stiffness for an element which undergoes compression would result in an underestimation of its stress level and thus a slow convergence of the iterative scheme. Similarly, providing high compressive stiffness for the stretched element will result in the same issue. In order to avoid this nonconformity, the direction of the stiffness homogenisation tests is decided based on the homogenised stress value. For this, the hydrostatic part is extracted from the averaged mesoscopic RVE stresses, and its value is used to judge on the stress state of the macroscopic element. If the hydrostatic stress is positive, then two uniaxial tensile tests are done, if it is negative - two compression tests. The uniaxial tests are performed in the horizontal and vertical direction. Additionally to the uniaxial tests, one pure shear test is done.

### 2.5. Nested multi-scale approach

As discussed in Subsection 2.2, the boundary conditions for the RVEs are prescribed in terms of the macroscopic deformation gradient. The macroscopic displacements, however, are not known *a priori* and depend on the deformation process inside the RVEs. In the multi-scale approach adopted here, deformation-driven iterative procedures are applied, where an initial macroscopic deformation is imposed and a series of solve steps are performed until reaching convergence. Subsequently, both the macroscopic and mesoscopic loads can be incremented, and the next iteration starts.

The homogenised mesoscopic stress passed to the macro-scale serves both to judge on the convergence of iterations and also to update the displacements at the macro-scale. The overall macroscopic stiffness tensor, composed of the homogenised stiffness tensors of the underlying RVEs, is used at the macro-scale to compute the increment of displacements. The algorithm for the current multi-scale implementation, executed by every processor, is detailed in Algorithm 1.

aggregates are explicitly represented. The parallel computation scheme of the current implementation of the multi-scale model is illustrated in Fig. 7. At the beginning of the simulation, the

---

**Algorithm 1: Multi-scale algorithm for ASR-affected concrete**

    **for** *every integration point at the macro-scale* **do**
        Generate an RVE;

    **for** *every time step i* **do**
        Apply boundary conditions at the macro-scale;
        **for** *every RVE* **do**
            Compute ASR products expansion value;
            Impose expansion at the ASR sites;
        **while** *solution is not converged* **do**
            Solve the macro-scale problem;
            **for** *every RVE* **do**
                Collect deformation gradient from the macro-scale;
                Apply it as boundary conditions;
                **while** *finite elements are damaged* **do**
                      Solve the meso-scale problem;
                      Reduce material properties of the damaged elements
                        (SLA);
                Homogenise stress;
                Determine hydrostatic component of the homogenised stress;
                Homogenise the RVE stiffness;
            Assemble global stiffness matrix;
            Assemble macro-scale internal force;
            Check for convergence;
        Output results;

---

### 2.6. Parallel implementation

The current ASR multi-scale model was implemented into the open-source finite-element library Akantu (Cuba Ramos, 2017; Richart et al., 2015; LSMS, 2016). The coding developments were carried out under the premise that the meso-scale computations for a single RVE are executed in serial. Therefore, the FE meshes of the RVEs must be small in size. This is the case for two-dimensional numerical concrete samples, in which only the coarse macro-scale FE mesh is partitioned among all available processors. Each processor generates an RVE for each integration point belonging to macroscopic finite elements.

### 3. Material behaviour at the meso-scale

It was previously reported that models with isotropic damage law without stiffness recovery in compression can capture macroscopic expansion of concrete under free boundary conditions and moderate uniaxial loading but fail under high compression (Dunant and Scrivener, 2012). Because the final goal of this study is to model a macro-structure with a mainly compressive stress state, such capacity is crucial. For that reason, we have adopted the fixed crack concept with stiffness recovery upon crack closure (Blaauwendraad et al., 1989). In this approach, one models continuum as an elastic-brittle orthotropic material with elastic properties degraded in a single direction. This direction is chosen at the first damaging event and kept fixed for the rest of the simulation (see Fig. 8). When a previously opened crack closes, it can again transmit stresses in the perpendicular direction. This capacity is important for complex stress states that develop in time. The direction of the crack is decided based on the principal stress criteria:

$$\sigma_i \leqslant f_t, \text{ with } i = 1, 2, 3,\qquad(23)$$

where $\sigma_i$ is the $i$th principal stress. If the stress $\sigma_i$ at any integration point violates Eq. 23, the plane perpendicular to the direction of $\sigma_i$ becomes the crack plane.

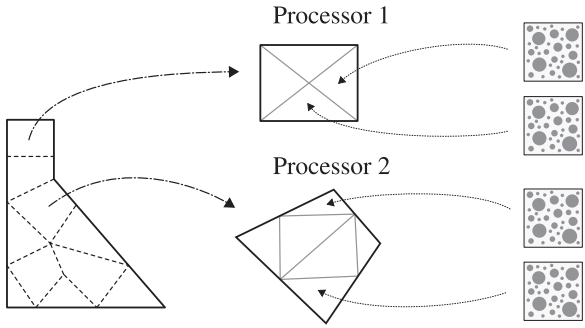

**Fig. 7.** Parallel computation scheme for multi-scale ASR simulations: the macroscopic finite-element mesh is partitioned among all available processors. Each processor generates the RVEs for the elements it owns. The RVEs are then used for the meso-scale computations.

Having a single crack passing through an element allows a simple energy accounting, in a way proposed by Bazant et al. (1983). The essence of this model is to smear out existing micro-cracks over the fracture process zone of width $w_c$, taken equal to the average element size. The area under the stress-strain curve in Fig. 5b equals the fracture energy $G_c$ divided by the crack band width $w_c$:

$$\frac{G_c}{w_c} = \int_0^\infty \sigma d\varepsilon^f = \frac{1}{2}f_t^2\left(\frac{1}{E} - \frac{1}{E_t}\right),\qquad(24)$$

where $\varepsilon^f$ is the fracturing strain, $E$ is the initial material stiffness in the direction perpendicular to the crack, $E_t$ is the tangent to the tensile stress-strain softening curve. In all meso-scale simulations, the crack band width, $w_c$, is taken equal to the element size of 0.5 mm. The ultimate strain $\varepsilon_u$ becomes a function of the fracture energy, the crack band width, and the material strength in tension $f_t$:

$$\varepsilon_u = \frac{2G_c}{w_c f_t}.\qquad(25)$$

The stress-strain relation for the orthotropic material in a compact matrix form is

$$\boldsymbol{\sigma} = \mathbf{C}(\boldsymbol{\varepsilon} - \boldsymbol{\alpha}\Delta T),\qquad(26)$$

with $\mathbf{C}$ being the stiffness matrix in the coordinate system associated with the global problem, $\boldsymbol{\alpha}$ the thermal expansion coefficient, and $\Delta T$ the temperature increase. Multiple works studied thermal stress in the vicinity of a crack (Kit et al., 1977; Atkinson and Clements, 1977; Da Yu Tzou, 1990). Among other findings, they have shown that the temperature gradient can cause crack propagation. Factors that play a role in this process include material properties, boundary conditions (both mechanical and thermal), crack shape, its dimension and the level of insulation. Considering the complexity of the problem, its finer scale and the lack of knowledge of material and crack properties, currently, we ignore the thermo-mechanical aspects of crack growth and discard the second term on the rhs of Eq. 26. This is however a simplification that should be addressed in future work.

An orthotropic material has three perpendicular symmetry planes. Basis vectors of this material are perpendicular to the symmetry planes. These vectors could differ from the basis vectors of the global coordinate system. In such a case, the transformation from the coordinate system associated with the material symmetry planes to the global coordinate system in matrix form can be expressed as

$$\mathbf{C} = \mathbf{Q} \cdot \mathbf{C}^{(p)} \cdot \mathbf{Q}^T,\qquad(27)$$

where $\mathbf{Q}$ is the transformation matrix and $\mathbf{C}^{(p)}$ is the orthotropic stiffness matrix in the direction of the basis vectors. In the scope of the crack model, matrix $\mathbf{Q}$ accounts for the angle $\alpha$ between the global coordinate system and the crack plane shown in Fig. 8. The stiffness matrix takes the form (Bower, 2009):

$$\mathbf{C}^{(p)} = \begin{bmatrix} c_{11} & c_{12} & c_{13} & 0 & 0 & 0 \\ & c_{22} & c_{23} & 0 & 0 & 0 \\ & & c_{33} & 0 & 0 & 0 \\ & \text{sym} & & c_{44} & 0 & 0 \\ & & & & c_{55} & 0 \\ & & & & & c_{66} \end{bmatrix}.\qquad(28)$$

Its components are related to the elastic constants:

$$c_{11} = E_1(1 - \nu_{23}\nu_{32})\gamma, \quad c_{22} = E_2(1 - \nu_{13}\nu_{31})\gamma, \quad c_{33} = E_3(1 - \nu_{12}\nu_{21})\gamma,$$
$$c_{12} = E_1(\nu_{21} + \nu_{31}\nu_{23})\gamma, \quad c_{13} = E_1(\nu_{31} + \nu_{21}\nu_{32})\gamma, \quad c_{23} = E_2(\nu_{32} + \nu_{12}\nu_{31})\gamma,$$
$$c_{44} = \mu_{23}, \quad c_{55} = \mu_{13}, \quad c_{66} = \mu_{12},$$
$$\gamma = 1/(1 - \nu_{12}\nu_{21} - \nu_{23}\nu_{32} - \nu_{31}\nu_{13} - 2\nu_{21}\nu_{32}\nu_{13}),$$

where $E_k$ is the Young's modulus in the direction of the basis vector $\mathbf{n}_k$, $\nu_{kl}$ and $\mu_{kl}$ are the Poisson's ratio and the shear modulus between directions $\mathbf{n}_k$ and $\mathbf{n}_l$. Symmetry of the stiffness matrix is ensured by satisfying the following equality:

$$\nu_{kl}/E_k = \nu_{lk}/E_l \text{ for } k,l = 1,2,3 \text{ and } k \neq l \text{ (no sums)}.\qquad(29)$$

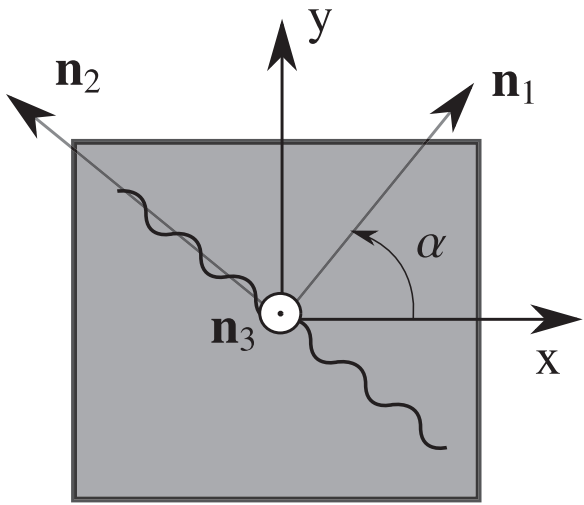

**Fig. 8.** Fixed crack model. Orientation axes of a crack $\mathbf{n}_1$ and $\mathbf{n}_2$ are turned with respect to the global coordinate system by an angle $\alpha$.

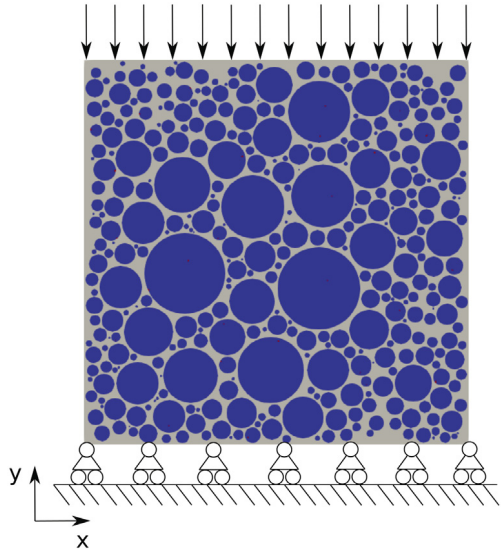

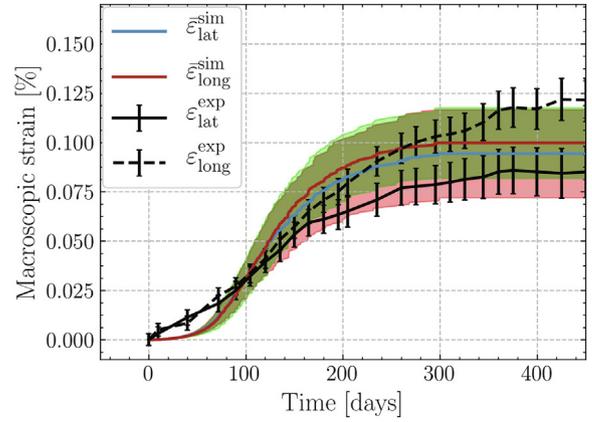

**Fig. 9.** Concrete meso-structure and boundary conditions for the calibration tests. Specimens are composed of aggregates (in blue) surround by mortar (in grey). The load is applied on top of the specimen during the compression test only.

**Table 1**
Material properties.

|  | $E$[GPs] | $\mu$[GPa] | $\nu$[−] | $G_c$ [J/m²] | $f_t^0$ [MPa] |
|---|---|---|---|---|---|
| Aggregates | 59 [1] | 22.6 | 0.3 [1] | 160 [2] | 10 [2] |
| Mortar | 12 [1] | 4.6 | 0.3 [1] | 60 [3] | 3 [1] |
| ASR product | 11 [4] | 4.7 | 0.18 [4] | – | – |

[1]Taken from Dunant and Scrivener (2012).
[2]Taken from Ben Haha (2006).[3]Taken from Xu et al., (2009).
[4]Taken from Leemann et al. (2013).

For the undamaged material, all the elastic properties equal the initial isotropic values:

$$E_k = E, \quad \nu_{kl} = \nu, \quad \mu_{kl} = \mu. \tag{30}$$

If a crack starts growing along the plane $\mathbf{n_2 n_3}$, as shown in Fig. 8, the material properties for the perpendicular direction $\mathbf{n_1}$ are reduced according to:

$$E_1 = E(1-d), \quad \nu_{12} = \nu_{13} = \nu(1-d), \quad \mu_{12} = \mu_{13} = \mu(1-d), \tag{31}$$

where $d$ is a damage parameter that was defined earlier. Material properties in the undamaged directions stay unchanged and equal to the initial isotropic value.

Modelling orthotropic damage facilitates the efficiency of accounting for stiffness recovery upon crack closure. Initially, the principal stress in the direction perpendicular to the crack is controlled. A negative value indicates that the crack segment is closed and its stiffness has to be recovered. In this study, we adopt frictionless crack approach and fully recover Young's moduli and Poisson's ratios while keeping the values of shear moduli reduced:

$$E_k = E, \quad \nu_{kl} = \nu, \quad \mu_{kl} = \mu(1-d).$$

The frictionless crack approach imposes a limitation on the ability of the model to reproduce real concrete behaviour under shear load, however, it was adopted in order to avoid convergence issues.

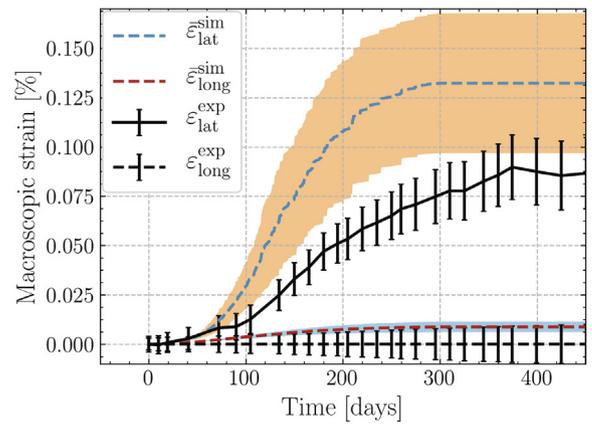

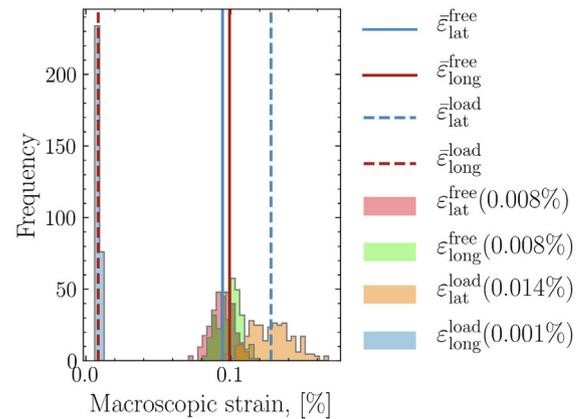

**Fig. 10.** Macroscopic strain of numerical concrete specimens affected by ASR under (a) free expansion conditions, (b) uniaxial compression loading of 10 MPa compared to the experimental results (Multon and Toutlemonde, 2006). Solid lines denote the average values, and the shaded areas cover the range of output parameter. (c) Distributions of the final expansion from the Monte Carlo simulations. Values in brackets are the standard deviations.

## 4. Meso-scale model calibration

While the elastic and fracture properties of concrete are available in the literature, the parameters of the ASR expansion law

(Eq. 10) require calibration with experiments. Values to be calibrated are the number of ASR sites, the asymptotic strain $\varepsilon(\infty)$, and the latency, $\tau_{lat}$, and characteristic, $\tau_{ch}$, times. Calibration is done based on the experimental results of Multon and Toutlemonde (2006), who studied the effect of loading on the macroscopic behaviour of ASR-affected concrete. For this purpose, cylindrical specimens of concrete with ASR were either free to expand or loaded longitudinally and radially. Control cylinders without ASR were set under the same conditions. Their deformations were taken out from similar measurements of the ASR specimens. It allowed separating the ASR-caused strain values from the concrete shrinkage and creep.

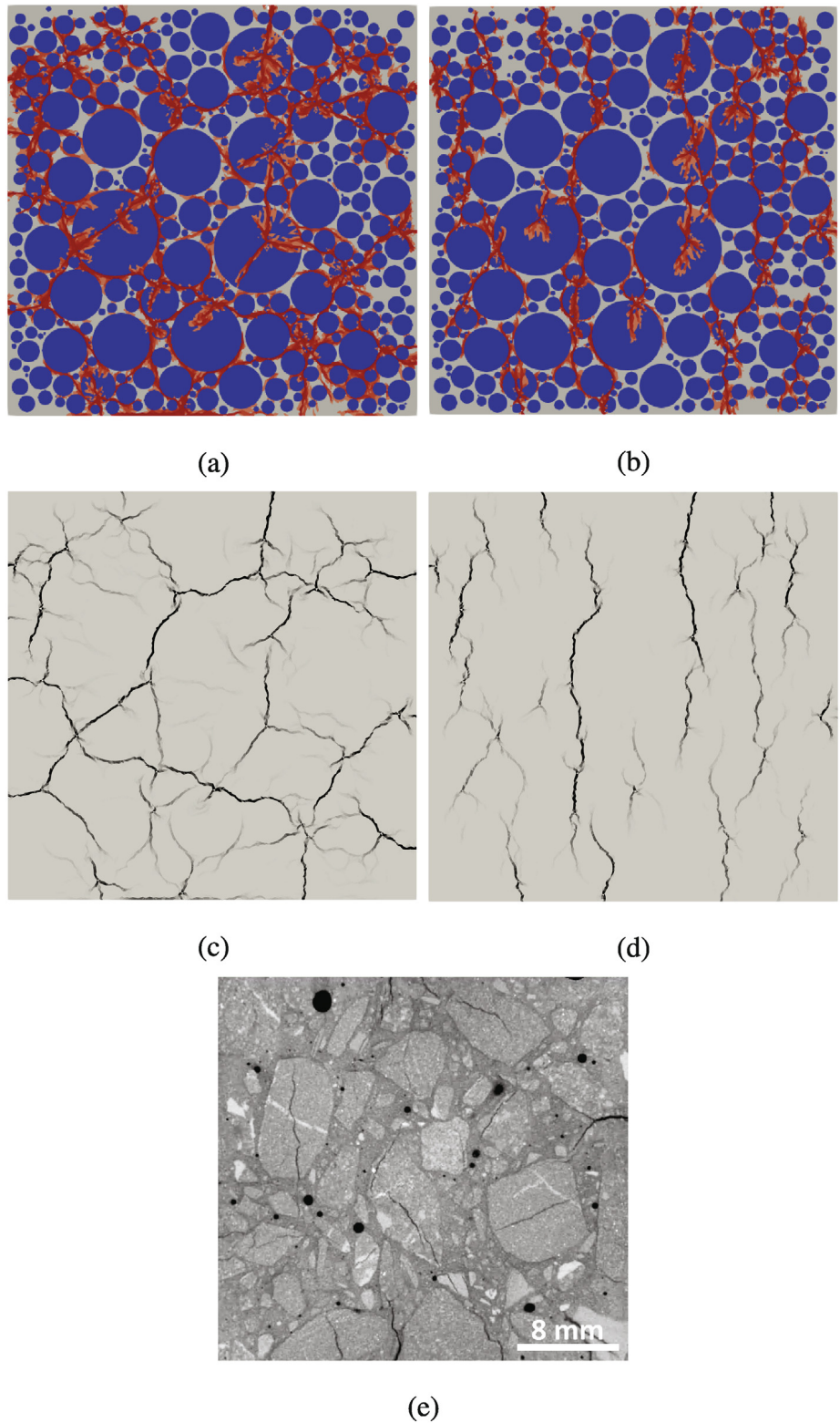

**Fig. 11.** Damage map within concrete specimens affected by ASR under (a) free expansion conditions, (b) uniaxial compression loading of 10 MPa. Red colour denotes fully damaged elements ($d = 1$). (c,d) Crack opening maps of the same specimens under the same conditions. (e) X-ray micro-tomography image of laboratory-based ASR-accelerated concrete under free expansion conditions (courtesy of the Concrete Technology Group at Empa [Swiss Federal Laboratories for Material Testing and Research, Dübendorf, Switzerland]).

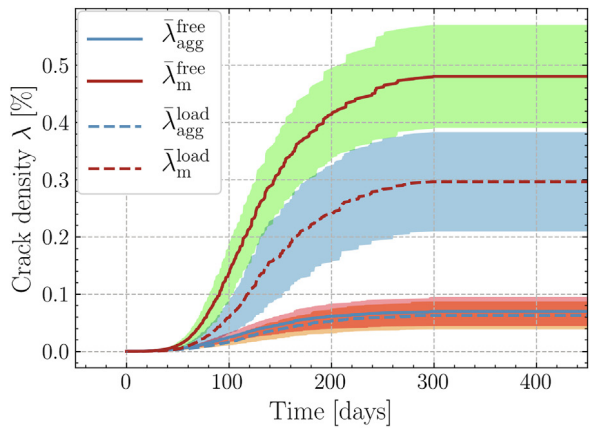

(a)

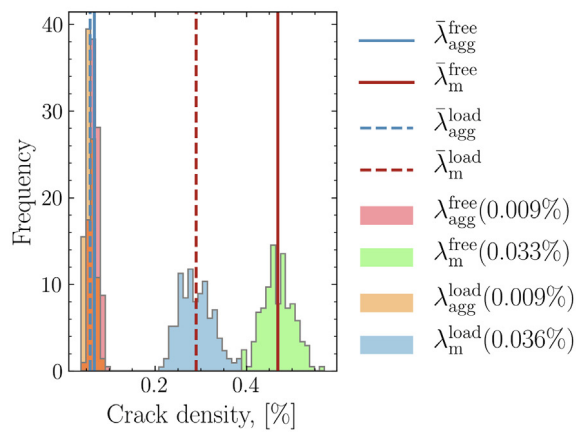

(b)

**Fig. 12.** (a) Crack density within aggregates and mortar for the free expansion and the loaded ASR simulations. (b) Distributions of the final crack density per phase with its mean values and standard deviations (values in brackets) from the Monte Carlo simulations. The free specimen has larger volume of cracks both in aggregates and mortar due to their random orientation.

The algorithm proposed by Wriggers et al. (2006) is used to generate the geometrical models of the concrete specimens and RVEs. In this algorithm, a circular shape is assumed for all aggregates and the Fuller curve is used as a grading curve for the generation of aggregates. We are not, however, restricted by the shape of the aggregates. Any other aggregate shapes can be incorporated into the proposed model with ease. Subsequently, the geometrical models are discretised with Gmsh (Geuzaine et al., 2009) into finite-element meshes of linear triangular elements with a uniform average element size $h = 0.5$mm. The resulting numerical concrete samples have a size of $70 \times 70$mm$^2$, and contain aggregates of circular shape with diameters in the range $1 - 16$mm (see Fig. 9). Concrete specimens were generated with an aggregate packing density of 70%.

Mechanics of the ASR product accounts for different phenomena taking place in a range of scales varying from nanometre to millimetre, and not yet fully understood. These processes include the growth of ASR product (either amorphous or crystalline), the opening of micro-cracks due to that, possible transport of product into pores and fissures, change of its properties with time and the surrounding environment. Leemann and collaborators (Leemann et al., 2016; Leemann et al., 2019) have shown that the primary ASR product starts to accumulate between mineral grains within reactive concrete aggregates. Therefore, a typical size of an ASR inclusion is few nanometres. In the current numerical study, the size of an element is 0.5 mm. Thus, the expansion that we apply at a single finite element to represent the effect of ASR, should be seen as a homogenised expansion of a portion of an aggregate surrounding a single pocket of the ASR product, rather than the expansion of the product itself. In order to reproduce the ASR sites, a certain number of finite elements are randomly chosen inside the aggregates and assigned the mechanical properties of the ASR product.

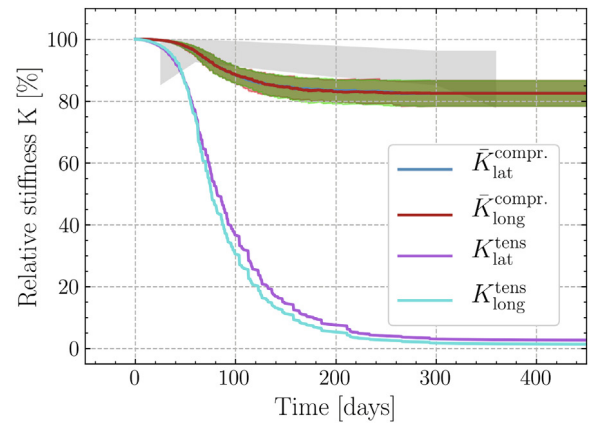

(a)

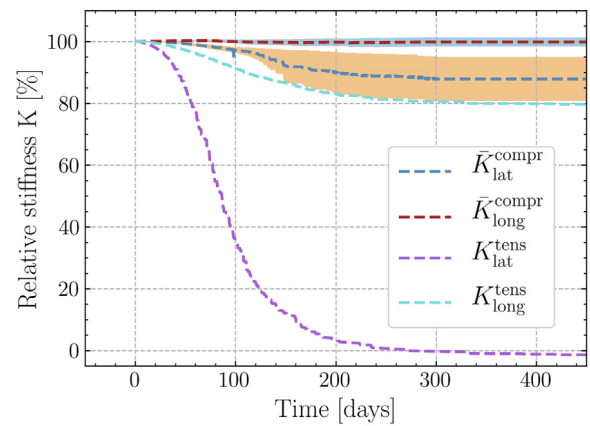

(b)

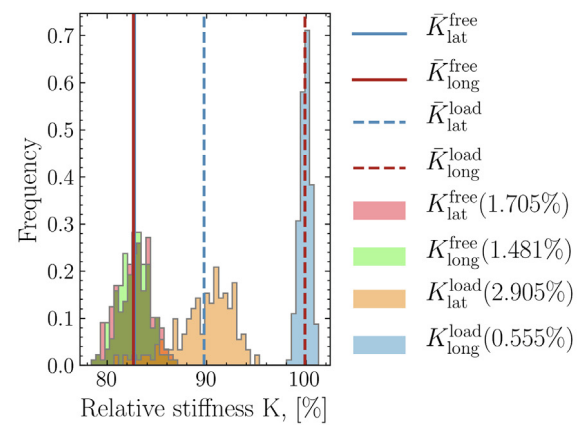

(c)

**Fig. 13.** Estimation of the stiffness loss via compressive and tensile tests under (a) free expansion conditions, (b) uniaxial loading of 10 MPa. The grey shaded area corresponds to the range of experimental values reported by Ben Haha)2006. (c) Distributions of the final relative stiffness in two directions with their mean values and standard deviations (values in brackets) from the Monte Carlo simulations.

**Table 2**
Parameters of the ASR-expansion law.

| | |
|---|---|
| Ratio of the ASR sites area to the aggregates area | 0.1 % |
| Asymptotic strain, $\varepsilon(\infty)$ | 6.5 % |
| Latency time, $\tau_{lat}(T_0 = 38°C)$ | 30 days |
| Characteristic time, $\tau_{ch}(T_0 = 38°C)$ | 60 days |
| Thermal activation constant, $U_C$ | 5,400 K |
| Thermal activation constant, $U_L$ | 9,700 K |

The heterogeneity of the underlying micro-structure of the aggregates and the mortar is taken into account by locally varying the tensile strength of the corresponding finite elements according to a Weibull distribution. The cumulative distribution of the tensile strength $f_t$ is computed as follows:

$$F(f_t) = 1 - \exp\left(-\left(\frac{f_t - 0.8 f_t^0}{\lambda}\right)^k\right), \quad (32)$$

where $k$ and $\lambda$ are the shape and the scale factors of the Weibull distribution, taken equal to 5 and $0.2 \cdot f_t^0$ correspondingly.

The components of concrete specimens have the material properties listed in Table 1. Shear modulus $\mu$ is computed as $\mu = E/[2(1 + \nu)]$.

The boundary conditions for the calibration tests are shown in Fig. 9. Two simulation types are performed: the first one being the ASR-induced free expansion and the second one the loaded expansion. In the first test type, only the internal load coming from the expanding ASR-sites is present. In the second one, a vertical compression load of 10 MPa is additionally applied on top. Similar to the experiments, both simulation types last 450 days. The time step is taken to be equal to 0.5 days. This value provides a good time resolution and allows to capture the gradual evolution of the crack network.

A single-simulation outcome strongly depends on the specific realization of a concrete specimen (aggregates sizes and positions, spatial distribution of tensile strength, positions of ASR-sites) and may not be representative. To obtain a statistical distribution of results, a Monte Carlo simulation consisting of 200 independent runs is performed for each loading type. The random parameter of the Monte Carlo analysis is the concrete meso-structure. Out of these 200 runs, one hundred has a fixed mesh but different distribution of ASR-sites (their density is always kept unchanged). Second half of the Monte Carlo runs are performed with the different spatial distribution of aggregates and consequently ASR-sites too. The statistics of output parameters, including mean values (letters with overline) and standard deviations (percent in brackets) are plotted in Fig. 10-12 and 13.

The parameters of the ASR-expansion law and the density of the ASR sites were calibrated on the free expansion experiment and kept the same for the loaded case and the simulations presented in the following section. They are listed in Table 2. The total number of ASR sites (or the ratio of their total area to the total area of aggregates) is an important parameter, which governs crack distribution across the domain: a small number of sites results in more isolated cracks, while high number leads to a more diffused crack pattern. Currently, quantifying this parameter is a complicated task as there is no unique way of accurately determining the reactive sites. In our simulations, this parameter was chosen in a rather intuitive way.

Results of the calibration tests and the experimental expansion values are plotted in Fig. 10. Set of macroscopic expansion curves of the numerical specimens is compared to the experimental ones. The latter represent the average expansion values over multiple tested specimens. The range of the observed values is marked by the error bars. Both experimental and numerical curves account only for the ASR-caused expansion: deformation of concrete due to its shrinkage and the load application were previously taken out.

In Fig. 10a, the numerical expansion curves for the freely expanding specimens are compared the experimental ones. Both sets of curves have typical sigmoidal shapes and similar asymptotic values. The main discrepancies between the two sets of curves are the faster experimental expansion rate during the first 50 days and the earlier saturation of the numerical expansion (300 days against ~370 days in the experiment). To improve the fitting, we could have used another thermal law, however, the choice was made in favour of the one presented in Eq. 10 as it is the one most commonly used in the field of ASR modelling.

In Fig. 10a, both the numerical and the experimental specimens show different strain values depending on the direction. While for the simulation the difference is insignificant, the experimental results show more pronounced separation. In the free-expansion simulation, cracks are overall randomly oriented, however they may have a slight orientation bias due to the aggregates shape and positioning or the ASR-sites placement (see Fig. 11a). In the real concrete, the difference in expansions may also be caused by the casting directions, the effect of the self-weight, specific boundary conditions, rock anisotropy, etc. (Larive, 1998; Ben Haha, 2006; Gautam et al., 2016). Macroscopic strain curves for both free expansion and loaded conditions have a typical S-shape, as a result of the similar shape of the ASR-product expansion curve. It is important to mention, that in the field conditions, the ASR-affected dams experience continuous swelling during decades (Mauris et al., 2015). The difference in long-term behaviour of the laboratory specimens with ASR and the full-scale structures is subject of debate. One of the possible cause of the expansion halt is the alkalis concentration drop due to its leaching from specimens.

The numerical prediction of the loaded specimen shows also similar trends as the experimental one (see Fig. 10b). Both in the experiment and in the simulation, application of 10 MPa load reduces significantly the concrete expansion along the loading direction. The experimental average longitudinal expansion, $\varepsilon_{\text{long}}^{\text{exp}}$, is almost null. The numerical average $\bar{\varepsilon}_{\text{long}}^{\text{sim}} = 0.009\%$. The experimental strain in the stress-free direction, $\varepsilon_{\text{lat}}^{\text{exp}}$, is slightly bigger than the average values of the free expansion case, which is also reproduced in the simulations. However, the numerical prediction, $\bar{\varepsilon}_{\text{lat}}^{\text{sim}}$, overestimates this increase in the lateral expansion and has a wider distribution of the final values. This is due to the effect of the concrete meso-structure, the material properties variation, and the specific distribution of the ASR sites. Using the real concrete tomography to build a numerical model and then comparing two sets of results could improve the model's accuracy. Another source of difference could be the 2D nature of the model: cracks are more free to spread in a 3D volume, which in its turn reduces overall expansion. Nonetheless, the clear advantage of this model is its ability to have positive longitudinal expansion under high-load conditions due to the orthotropic damage model with stiffness recovery, which was not the case in the previous model of Dunant et al. (2012).

It is important to mention that the final expansion values shown in Fig. 10c have distribution close to normal. Similar shapes could be also appreciated in Fig. 12b-13c. Such univariate distribu-tion suggests that varying positions of ASR-sites within fixed aggregates and shuffling aggregates themselves has a similar effect on the output results.

The damage patterns are shown in Fig. 11a-11b. Cracks, which were earlier opened but later closed and recovered normal stiffness, do still appear in red. This leads to a certain visual overestimation of the damage amount. Instead of the damage parameter $d$, Fig. 11c-11d show the additional area of damaged elements. Basic assumption of this plotting way is that, in a damaged element, most of the deformation comes from the crack opening. Then, the crack area of a finite element, $A^{cr}$ could be computed as follows

$$A^{cr} = A_0^{el} \cdot \varepsilon_{vol}, \qquad (33)$$

where $\varepsilon_{vol}$ is the volumetric strain, $A_0^{el}$ is the initial area of an element. In 3D-computations, an analogue of the crack area would be the crack volume. This metric is also facilitating comparison with the experiments, where crack density is one of the output parameters (it will be discussed later in this section). Comparing two sets of figures we could note that some of the cracks, which are explicit in the damage maps, could not be distinguished in the crack opening plots. This is the effect of the crack closure and stiffness recovery, which is happening due to the stress field of one crack being disturbed by the approaching of another one.

Results of the numerical simulation are compared to the real crack pattern coming from the laboratory experiments, plotted in Fig. 11e. The image is obtained by X-ray micro-tomography of the ASR-accelerated concrete under free-expansion conditions after 500 days. This cracking pattern is the result of storing an ASR-affected concrete block inside alkaline solution under constant temperature of $40°C$. Similar to the numerical results, real concrete cracks are present both in aggregates, paste and interfacial transition zone. Their random orientations, broken shapes and equal spread across the cross-section resemble those from Fig. 11c.

The response of the meso-structure to the load application shows cracks alignment with the load direction (Fig. 11b-11d). The crack pattern explains why the expansion in the horizontal is higher than the one in the vertical direction (Fig. 10b). Since multiple coalesced vertical cracks are widely opened in the horizontal direction, the corresponding macroscopic strain, $\varepsilon_{lat}^{num}$, is larger than the strain in the direction of the load, $\varepsilon_{long}^{num}$.

Fig. 12 plots the crack density, $\lambda$, which is the ratio between the crack area within a specific phase and the total area of this phase. To compute this parameter, the crack area from Eq. 33 is used. Resulting crack density per phase is not exceeding one percent. Recent tomography studies of the specimens in the laboratory and field conditions showed total crack densities of a similar order (total density of $1.8 - 2.6\%$) (Shakoorioskooie et al., 2019; Leemann et al., 2019).

For both simulations, we can observe smaller crack density in aggregates compared to the mortar caused by the lower strength of the latter. Previous studies on the effect of creep on ASR show that it helps to accommodate larger strains in the cement paste without damaging it (Giorla et al., 2015; Rezakhani et al., 2020). An applied model combining isotropic sequential linear analysis with visco-elasticity was proposed by Gallyamov et al. (2021). The current orthotropic model does not include visco-elastic effects, thus slightly overestimates the damage amount in the mortar.

Another interesting observation from the same plot concerns the crack density within aggregates. Both under the free-expansion and loaded conditions, it has similar average value, which suggests that there is a very weak effect of the macroscopic load on the damaging of aggregates due to the ASR. Considering Fig. 10a-10b and Fig. 12a together, we could conclude that the major part of the macroscopic expansion is due to the damage in the mortar rather than the aggregates.

The stiffness loss is plotted in Fig. 13. The relative stiffness, expressed in per cent, is computed by applying either compressing or extending displacements on one specimen's surfaces. After solving for a new displacement field, internal forces at the nodes along this surface are integrated into a scalar. The ratio between the current value of this integral and its initial value serves as an estimator of the specimen's stiffness. Pulling or squeezing a sound piece of concrete would give rise to high reaction forces, whereas in a severely damaged specimen the applied load will only open cracks. For the Monte Carlo simulations, stiffness was evaluated via compression, therefore the whole distribution is obtained. For the comparison purpose, a single simulation with stiffness computed via tension was additionally performed.

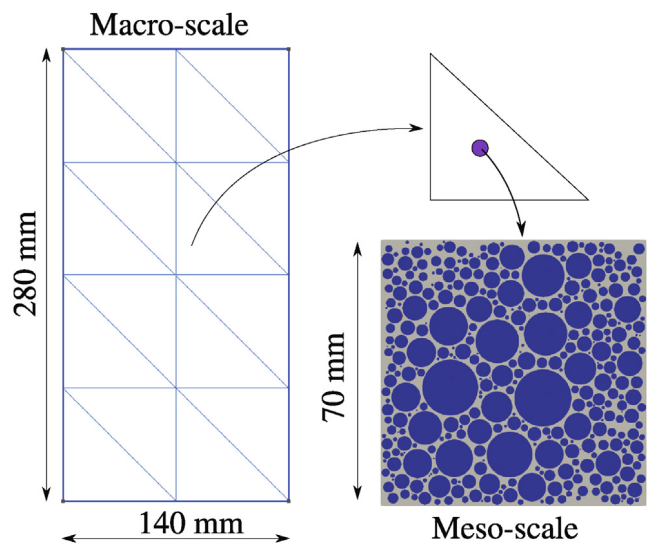

**Fig. 14.** Meshes used for the multi-scale simulation. Each of 16 macroscopic finite elements has a square concrete RVE at the meso-scale.

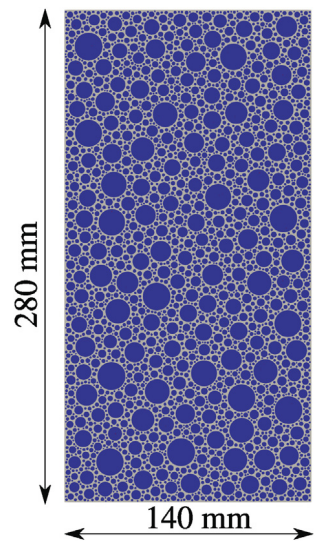

**Fig. 15.** Mesh for the detailed macro-scale simulation used for the comparison with FE$^2$.

From both tests, we see the relative stiffness estimated in tension being lower than the compressive one. This is due to the stiffness recovery of cracks brought in contact. Although this phenomenon could still happen while applying tension, its effect is more pronounced during the compression test. Another interesting observation concerns the ratio between stiffness values into two directions. In the free expansion case the stiffness is almost equally reduced in both directions, whereas the loaded case shows larger stiffness loss in the x-direction. This anisotropy is equally pronounced in both tensile and compressive estimates and is linked to the vertical cracks percolation. The difference between stiffness values estimated via tensile and compressive tests justifies adaptive homogenisation procedure, where the loading type of tests is chosen based on the stress state of the macroscopic element. In Fig. 13a, the grey shaded area denotes the limiting values of stiffness in the accelerated free expansion experiments of Ben Haha (2006). The results of the current model show stiffness reduction of $10-20\%$, which is very close to the experimental values. As Cuba Ramos et al. (2018) have previously suggested, adding the crack interlock and closure effects indeed increases the load-bearing capacity of the numerical sample. This property of the meso-scale concrete behaviour permits integrating the proposed model into the multi-scale numerical model, where a single macroscopic element could transition from tension into compression and vice versa.

## 5. Multi-scale model verification

With a calibrated meso-scale ASR model, the multi-scale model can now be tested. To the best of these authors' knowledge, there are no sets of laboratory experiments and long-term field data on the same ASR-affected concrete available in the literature. To overcome this issue, we propose using the same laboratory experiments by Multon and Toutlemonde (2006) as a macro-scale structure to test the ASR multi-scale model and to provide the proof of concept for the method. Although the separation of scales concept does not hold anymore, this choice allows a direct comparison between the results of the $FE^2$-simulation and the experimental measurements. Another advantage of this set up is its simple stress state, which is not a case for a large operating structure such as a dam or a bridge pier. By applying a load at the macro-scale, we can directly observe how it is balanced by the internal forces at the meso-scale.

Fig. 14 shows the multi-scale set up used for the validation test. The size of the numerical macro-scale specimen is $140 \times 280$ mm$^2$, while the RVE size is kept the same as the one used previously ($70 \times 70$ mm$^2$). In the present case, each finite element at the macro-scale is linked to an underlying RVE, which represents the heterogeneous structure of concrete at the meso-scale. The total number of RVEs corresponds to the number of finite elements in the macroscopic FE mesh because each macroscopic finite element contains only one integration point. The macro-scale mesh is coarsely discretized for the consistency reason. Namely, in order to keep the macro-element size larger or equal to the RVE size in partial accordance with the separation of scales concept. From the computational point of view, the macro-scale problem could be easily discretized by a much denser mesh.

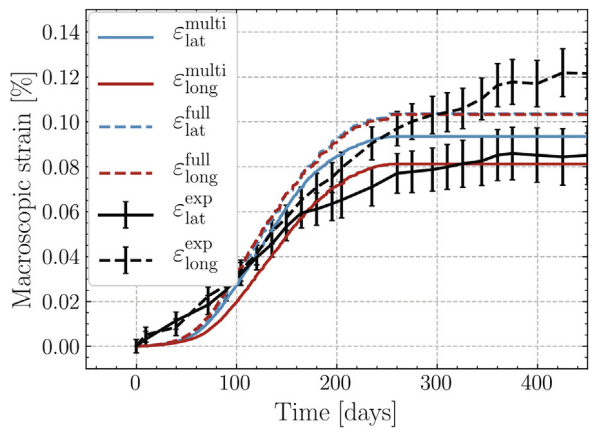

(a)

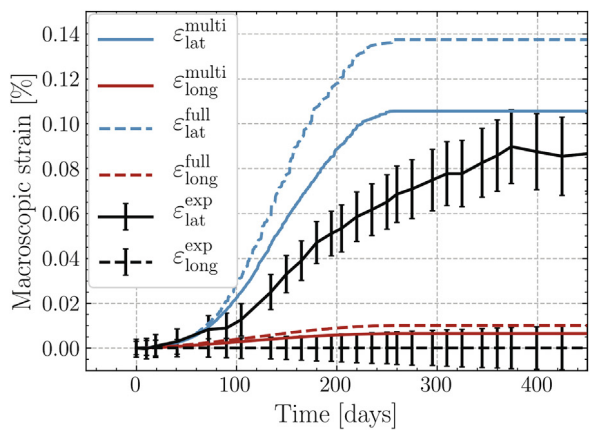

(b)

**Fig. 16.** Expansion curves of the macroscopic specimen obtained by multi-scale and detailed macro-scale simulations: (a) free expansion experiment; (b) uniaxial compression by 10 MPa. The average values of the experimental expansion curves are plotted for comparison. The two sets of numerical results are in a good agreement and are also close to the experimental values.

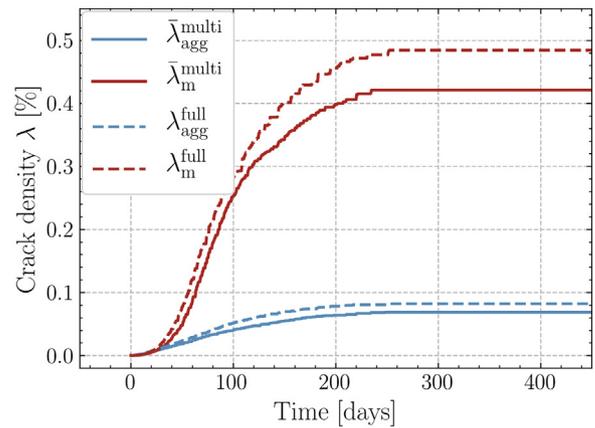

(a)

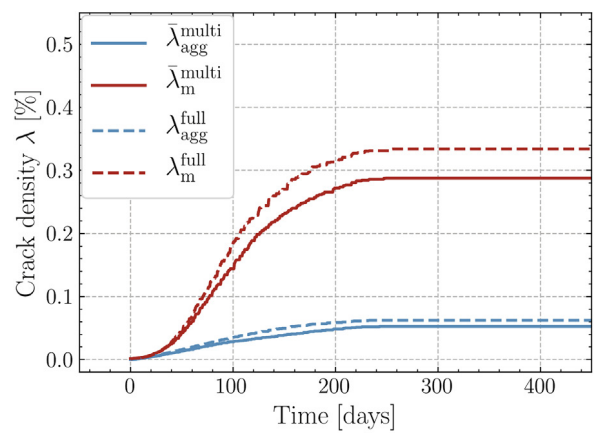

(b)

**Fig. 17.** Average crack density for the (a)free and (b)loaded multi-scale simulation.

Laboratory experiments were performed at 38°C temperature (Multon, 2004). Since the temperature field is homogeneous within the specimen, a constant single value $T_M$ is passed to all RVEs during all time steps.

For comparison, fully detailed simulations of the macro-scale structure were performed. The mesh for these simulations is shown in Fig. 15. In both models, the density of the ASR sites and the parameters of the ASR product expansion law from the calibra-tion tests is used (see Table 2). The two following tests were performed: free expansion and uniaxial compression with 10 MPa. This proposed stress range is chosen to simulate possible stress variations within a massive concrete structure undergoing ASR. As the statistical distributions of the output parameters were studied in the previous section, here, we limit ourselves to a single realisation of both the multi-scale and the fully detailed models.

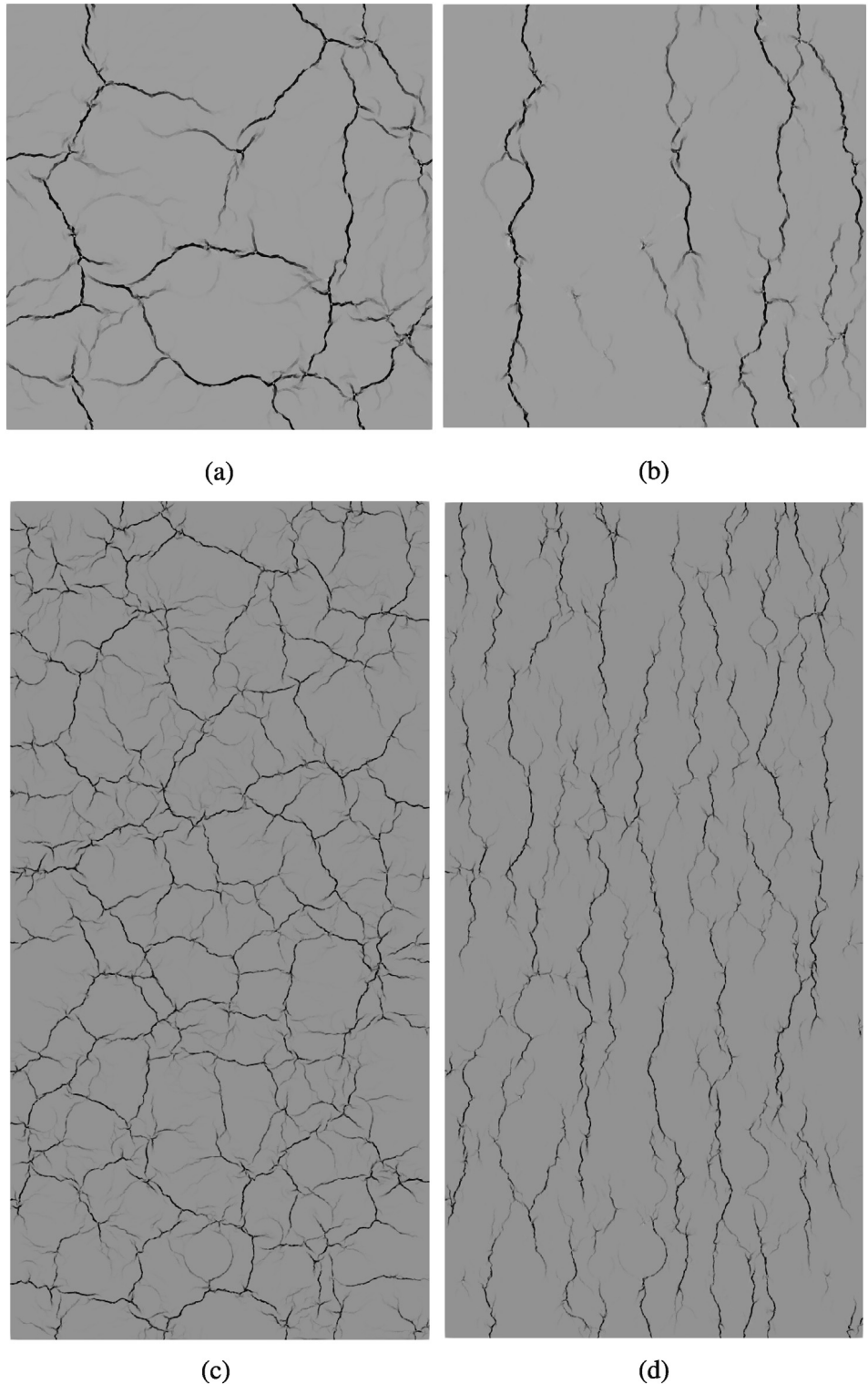

**Fig. 18.** Crack opening patterns within typical RVEs under (a)the free expansion conditions and (b)the uniaxial compression stress of 10 MPa. (c,d) Similar images for the full macro-scale model. Crack patterns and their amounts are in fair agreement between the two sets of simulations.

Expansion curves from the multi-scale simulations are plotted in Fig. 16 next to the ones coming from the fully detailed macro-scale simulations. Expansion values were obtained by averaging the strain values at the macro-level. We see a good match between these two sets of curves in both simulations. Naturally some differences remain between multi-scale and detailed macro-scale simulations. Similar to the calibration simulations, there is a gap between the numerical lateral expansion value and its experimental counterpart. All these variations are because of the specific modelling of the detailed macro-scale model, which is not exactly reproduced by the multi-scale model.

The average crack density of the multi-scale simulations is plotted in Fig. 17, where averaging was done over all RVEs. The $FE^2$ simulation of the 10 MPa compression has the lowest crack area both in aggregates and mortar (0.05% and 0.29% correspondingly). It is followed by the free expansion experiment with 0.07% of aggregates and 0.42% of mortar occupied by cracks.

Crack patterns within RVEs are shown in Figs. 18a-18b. For comparison, similar images in the detailed macro-scale simulations are given in Fig. 18c-18d. A similar amount of cracks and their orientation in each pair of simulations is evident. The difference in cracks volume between the two loading cases is somehow expected. Compressive load pushes the stress state of each finite element further from the tensile failure envelop. In such a case, cracks are strongly localised and follow the direction perpendicular to the maximum principal stress. In case of a free expansion experiment, the stress state is closer to the limits, therefore cracks appear earlier and grow longer. Periodic boundary conditions applied at the RVEs cause cracks continuity across the borders. Artificial cracks perpendicularity to the periodic borders is a well-known issue, which also brings bias into the homogenised stiffness tensor. A possible solution was proposed by Coenen et al. (2012), where the boundaries align with the evolving localisation bands, enabling crack bands to grow slantwise to the boundaries directions. Potentially, this method could be adopted in the current model and enhance crack patterns and stiffness estimation.

## 6. Conclusions

A multi-scale finite-element method for simulating the mechanical consequences of ASR in large concrete structures has been presented. The multi-scale approach is advantageous because the macroscopic material behaviour in the simulation is governed by the mesoscopic ASR phenomena. Consequently, no drastic assumptions on how the mesoscopic damage evolution affects the effective stiffness of the concrete are required. The multi-scale model has been implemented in the parallel open-source library Akantu. Meso-scale laboratory experiments of Multon and Toutlemonde (2006) are numerically simulated as proof of concept. The model is able to simultaneously account for the loading due to the boundary conditions on the macro-scale and for the loading due to the ASR expansion at the meso-scale. The results, furthermore, demonstrate that the macroscopic stress state influences the orientation of damage inside the underlying RVEs. The effective stiffness becomes anisotropic, in cases where damage is aligned inside the RVE. Comparison of the multi-scale and the fully detailed macro-scale model shows a fair agreement of results both for free and loaded expansion tests. Although the evolution of the percolated crack bands within RVEs is evident, their diffused patterns make the classical homogenisation concept still applicable.

Another important development introduced in this model is the mesoscopic material law, which allows for the orthotropic reduction of elastic properties upon crack nucleation and their recovery upon crack closure. It was shown that this constitutive law better reproduces trends in macroscopic expansion and stiffness loss observed in the laboratory experiments than the isotropic damage without contact.

The performed validation of the multi-scale model suggests its applicability for detailed studies of real concrete structures affected by ASR. Such a study of a dam is planned in future work, including accounting for a different reaction extent for each RVE due to a non-uniform thermal field in the dam. Combining the link between the temperature and the mesoscopic eigenstrain with the temperature gradients at the macro-scale due to variable boundary conditions (Comi et al., 2012) would result in a non-uniform internal loading within each RVE. This is expected to emphasize the anisotropy of the ASR-affected concrete, which in the first place is triggered by the boundary conditions.

## Declaration of Competing Interest

The authors declare that they have no known competing financial interests or personal relationships that could have appeared to influence the work reported in this paper.

## Acknowledgments

The authors would like to thank the financial support of the Swiss National Science Foundation within the Sinergia project "Alkali-silica reaction in concrete (ASR)", grant number CRSII5_17108.

## References

Alnaggar, M. Cusatis, G., Di Luzio, G., 2013. Lattice Discrete Particle Modeling (LDPM) of Alkali Silica Reaction (ASR) deterioration of concrete structures. Cem. Concr. Compos., ISSN 09589465. doi: 10.1016/j.cemconcomp.2013.04.015.

Atkinson, C., Clements, D.L., 1977. On some crack problems in anisotropic thermoelasticity. Int. J. Solids Struct., ISSN 00207683. doi: 10.1016/0020-7683(77)90071-3.

Bangert, F., Kuhl, D., Meschke, G., 2004. Chemo-hygro-mechanical modelling and numerical simulation of concrete deterioration caused by alkali-silica reaction. Int. J. Numer. Anal. Methods Geomech., 28 (7–8): 689–714. ISSN 03639061. doi: 10.1002/nag.375.

Bazant, Z.P., Oh, B., 1983. Crack band theory for fracture of concrete. Mater. Struct. Constr. 16, 155–177. https://doi.org/10.1007/BF02486267.

Bazant, Z.P., Zi, G., 2000. Fracture mechanics of ASR in concretes with waste glass particles of different sizes. J. Eng. Mech. 126 (February), 226–232.

Ben Haha, M., 2006. Mechanical effects of alkali silica reaction in concrete studied by sem-image analysis. PhD Thesis 3516, 1–232.

Blaauwendraad, J., Rots, J.G., 1989. Crack models for concrete: discrete or smeared? Fixed, multi-directional or rotating? Heron 34 (1), 59. 0046-7316.

Bower, A.F., 2009. Applied Mechanics of Solids. CRC Press, Boca Raton. https://doi.org/10.1201/9781439802489.

Capra, B., Bournazel, J.P., 1998. Modeling of induced mechanical effects of alkali-aggregate reactions. Cem. Concr. Res., 1998. ISSN 00088846. doi: 10.1016/S0008-8846(97)00261-5.

Capra, B., Sellier, A., 2003. Orthotropic modelling of alkali-aggregate reaction in concrete structures: Numerical simulations. Mech. Mater. https://doi.org/10.1016/S0167-6636(02)00209-0.

Coenen, E.W.C., Kouznetsova, V.G., Geers, M.G.D., 2012. Multi-scale continuous-discontinuous framework for computational- homogenization-localization. J. Mech. Phys. Solids. ISSN 00225096. doi: 10.1016/j.jmps.2012.04.002.

Comby-Peyrot, I., Bernard, F., Bouchard, P.O., Bay, F., Garcia-Diaz, E., 2009. Development and validation of a 3D computational tool to describe concrete behaviour at mesoscale. Application to the alkali-silica reaction. Comput. Mater. Sci. ISSN 09270256. doi: 10.1016/j.commatsci.2009.06.002.

Comi, C., Fedele, R., Perego. U., 2009. A chemo-thermo-damage model for the analysis of concrete dams affected by alkali-silica reaction. Mech. Mater., 41 (3): 210–230. ISSN 01676636. doi: 10.1016/j.mechmat.2008.10.010.

Comi, C., Kirchmayr, B., Pignatelli, R., 2012. Two-phase damage modeling of concrete affected by alkali-silica reaction under variable temperature and

humidity conditions. Int. J. Solids Struct., 49 (23–24): 3367–3380. ISSN 00207683. doi: 10.1016/j.ijsolstr.2012.07.015.

Computational Solid Mechanics Laboratory at Ecole Polytechnique Federale de Lausanne. Akantu User 's Guide, 2016. URL URL:https://www.epfl.ch/labs/lsms/high-performance-computing/software/akantu/.

Cuba Ramos, A.I., 2017. Multi-Scale Modeling of the Alkali-Silica Reaction in Concrete. PhD thesis, EPFL.

Cuba Ramos, A.I., Roux-Langlois, C., Dunant, C.F., Corrado, M., Molinari, J.F., 2018. HPC simulations of alkali-silica reaction-induced damage: Influence of alkali-silica gel properties. Cem. Concr. Res., 109 (April): 90–102. ISSN 00088846. doi: 10.1016/j.cemconres.2018.03.020.

Tzou, Da Yu, 1990. The singular behavior of the temperature gradient in the vicinity of a macrocrack tip. Int. J. Heat Mass Transf. ISSN 00179310. doi: 10.1016/0017-9310(90)90198-4.

DeJong, M.J., Hendriks, M.A., Rots, J.G., 2008. Sequentially linear analysis of fracture under non-proportional loading. Eng. Fract. Mech. ISSN 00137944. doi: 10.1016/j.engfracmech.2008.07.003.

Dunant, C.F., Bentz, E., 2015. Algorithmically imposed thermodynamic compliance for material models in mechanical simulations using the AIM method. Int. J. Numer. Methods Eng. 104, 963–982. https://doi.org/10.1002/nme.4969.

Dunant, C.F., Scrivener, K.L., 2010. Micro-mechanical modelling of alkali-silica-reaction-induced degradation using the AMIE framework. Cem. Concr. Res., 40 (4): 517–525. ISSN 00088846. doi: 10.1016/j.cemconres.2009.07.024.

Dunant, C.F., Scrivener, K.L., 2012. Effects of uniaxial stress on alkali-silica reaction induced expansion of concrete. Cem. Concr. Res., 42 (3): 567–576. ISSN 00088846. doi: 10.1016/j.cemconres.2011.12.004.

Dunant, C.F., Scrivener, K.L., 2012. Effects of aggregate size on alkali-silica-reaction induced expansion. Cem. Concr. Res. ISSN 00088846. doi: 10.1016/j.cemconres.2012.02.012.

Erinc, M., Van Dijk, M., Kouznetsova, V.G., 2013. Multiscale modeling of residual stresses in isotropic conductive adhesives with nano-particles. Comput. Mater. Sci. ISSN 09270256. doi: 10.1016/j.commatsci.2012.06.012.

Esposito, R., 2016. The Deteriorating Impact of Alkali-Silica Reaction on Concrete: Expansion and Mechanical Properties. PhD thesis, Delft University of Technology.

Fritzen, F., 2011. Microstructural modeling and computational homogenization of the physically linear and nonlinear constitutive behavior of micro-heterogeneous materials. ISBN 9783866446991. doi: 10.5445/KSP/1000023534.

Gallyamov, E., Rezakhani, R., Corrado, M., Molinari, J.F., 2021. Meso-scale modelling of ASR in concrete: effect of viscoelasticity. 16th Int. Conf. Alkali Aggreg. React. Concr.

Gautam, B.P., Panesar, D.K., 2016. A new method of applying long-term multiaxial stresses in concrete specimens undergoing ASR, and their triaxial expansions. Mater. Struct. Constr., 49 (9): 3495–3508. ISSN 13595997. doi: 10.1617/s11527-015-0734-z.

Geers, M.G.D., Kouznetsova, V.G., Brekelmans, W.A.M., 2010. Multi-scale computational homogenization: Trends and challenges. J. Comput. Appl. Math. https://doi.org/10.1016/j.cam.2009.08.077.

Geuzaine, C., Remacle, J.-F., 2009. Gmsh: A 3-D finite element mesh generator with built-in pre- and post-processing facilities. Int. J. Numer. Methods Eng. 79, 1309–1331.

Giorla, A.B., Scrivener, K.L., Dunant, C.F., 2015. Influence of visco-elasticity on the stress development induced by alkali-silica reaction. Cem. Concr. Res., 70: 1–8. ISSN 00088846. doi: 10.1016/j.cemconres.2014.09.006.

Grimal, E., Sellier, A., Multon, S., Le Pape, Y., Bourdarot, E., 2010. Concrete modelling for expertise of structures affected by alkali aggregate reaction. Cem. Concr. Res. ISSN 00088846. doi: 10.1016/j.cemconres.2009.09.007.

Hill, R., 1963. Elastic properties of reinforced solids: Some theoretical principles. J. Mech. Phys. Solids. ISSN 00225096. doi: 10.1016/0022-5096(63)90036-X.

Iskhakov, T., Timothy, J.J., Meschke, G., 2019. Expansion and deterioration of concrete due to ASR: Micromechanical modeling and analysis. Cem. Concr. Res., 115 (February 2018): 507–518. ISSN 00088846. doi: 10.1016/j.cemconres.2018.08.001. URL:https://doi.org/10.1016/j.cemconres.2018.08.001.

Kit, G.S., Nechaev, Y.K., Poberezhnyi, O.V., 1977. Determination of the stress intensity factors in a plate subjected to heat transfer. Sov. Appl. Mech., 13 (4): 365–369. ISSN 00385298. doi: 10.1007/BF00882935.

Kouznetsova, V.G., Brekelmans, W.A.M., Baaijens, F.P.T., 2001. Approach to micro-macro modeling of heterogeneous materials. Comput. Mech., 27 (1): 37–48. ISSN 01787675. doi: 10.1007/s004660000212.

Larive, C., 1998. Apports combinés de l'expérimentation et de la modélisation à la compréhension de l'alcali-réaction et de ses effets mécaniques. PhD thesis, Laboratoire Central des Ponts et Chaussées.

Leemann, A., Lura, P., 2013. E-modulus of the alkali-silica-reaction product determined by micro-indentation. Constr. Build. Mater., 44: 221–227. ISSN 09500618. doi: 10.1016/j.conbuildmat.2013.03.018.

Leemann, A., Munch, B., 2019. The addition of caesium to concrete with alkali-silica reaction: Implications on product identification and recognition of the reaction sequence. Cem. Concr. Res., 120: 27–35. ISSN 0008–8846. doi: 10.1016/J.CEMCONRES.2019.03.016.

Leemann, A., Katayama, T., Fernandes, I., Broekmans, M.A.T.M., 2016. Types of alkali–aggregate reactions and the products formed. Proc. Inst. Civ. Eng. - Constr. Mater., 169 (3): 128–135, jun 2016. ISSN 1747–650X. doi: 10.1680/jcoma.15.00059.

Leemann, A., Borchers, I., Shakoorioskooie, M., Griffa, M., Muller, C., Lura, P., 2019. Microstructural Analysis of ASR in Concrete - Accelerated Testing Versus Natural Exposure. In: Proc. Int. Conf. Sustain. Mater. Syst. Struct., pages 222–229.

Martin, R.P., Metalssi, O.O., Toutlemonde, F., 2012. Modelling of Concrete Structures Affected by Internal Swelling Reactions: Couplings Between Transfer Properties, Alkali Leachning and Expansion. 2nd Int. Conf. Microstruct. Relat. Durab. Cem. Compos.

Mauris, F., Martinot, F., Fabre, J.-P., Bourgey, P., Sausse, J., 2015. Synthesis of hydraulic structures behavior: lessons learned from monitored dams of EDF in France. In 3rd Int. Work. Long-Term Behav. Environ. Friendly Rehabil. Technol. Dams, number October, Nanjing, China.

Miehe, C., 1996. Numerical computation of algorithmic (consistent) tangent moduli in large-strain computational inelasticity. Comput. Methods Appl. Mech. Eng. ISSN 00457825. doi: 10.1016/0045-7825(96)01019-5.

Multon, S., 2004. Evaluation expérimentale et théorique des effets mécaniques de l'alcali-réaction sur des structures modèles. PhD thesis, Laboratoire central des ponts et chaussées.

Multon, S., Sellier, A., 2016. Multi-scale analysis of alkali-silica reaction (ASR): Impact of alkali leaching on scale effects affecting expansion tests. Cem. Concr. Res. ISSN 00088846. doi: 10.1016/j.cemconres.2015.12.007.

Multon, S., Toutlemonde, F., 2006. Effect of applied stresses on alkali-silica reaction-induced expansions. Cem. Concr. Res., 36 (5): 912–920. ISSN 00088846. doi: 10.1016/j.cemconres.2005.11.012.

Multon, S., Sellier, A., Cyr, M., 2009. Chemo-mechanical modeling for prediction of alkali silica reaction (ASR) expansion. Cem. Concr. Res. ISSN 00088846. doi: 10.1016/j.cemconres.2009.03.007.

Omikrine Metalssi, O., Seignol, J.F., Rigobert, S., Toutlemonde, F., 2014. Modeling the cracks opening-closing and possible remedial sawing operation of AAR-affected dams. Eng. Fail. Anal., 36: 199–214. ISSN 13506307. doi: 10.1016/j.engfailanal.2013.10.009.

Pignatelli, R., Comi, C., Monteiro, P.J., 2013. A coupled mechanical and chemical damage model for concrete affected by alkali-silica reaction. Cem. Concr. Res. ISSN 00088846. doi: 10.1016/j.cemconres.2013.06.011.

Puatatsananon, W., Saouma, V., 2013 . Chemo-mechanical micromodel for alkali-silica reaction. ACI Mater. J., 110 (1): 67–77. ISSN 0889325X. doi: 10.14359/51684367.

Rezakhani, R., Alnaggar, M., Cusatis, G., 2019. Multiscale Homogenization Analysis of Alkali-Silica Reaction (ASR) Effect in Concrete. Engineering. ISSN 20958099. doi: 10.1016/j.eng.2019.02.007.

Rezakhani, R., Gallyamov, E., Molinari, J.F., 2020. Meso-scale Finite Element Modeling of Alkali-Silica-Reaction. Manuscript submitted for publication.

Richart, N., Molinari, J.F., 2015. Implementation of a parallel finite-element library: Test case on a non-local continuum damage model. Finite Elem. Anal. Des., 100: 41–46. ISSN 0168–874X. doi: 10.1016/J.FINEL.2015.02.003.

Rots, J.G., 2001. Sequentially linear continuum model for concrete fracture. Fract. Mech. Concr. Struct. Proc. 4th Int. Conf. Fract. Mech. Concr. Struct., pages 831–839, 2001. ISSN 00223050. doi: 10.1136/jnnp-2014-307772.

Rots, J.G., Invernizzi, S., 2004. Regularized sequentially linear saw-tooth softening model. Int. J. Numer. Anal. Methods Geomech., 28 (7–8): 821–856. ISSN 03639061. doi: 10.1002/nag.371.

Rots, J.G., Belletti, B., Invernizzi, S., 2008. Robust modeling of RC structures with an event-by-event strategy. Eng. Fract. Mech. ISSN 00137944. doi: 10.1016/j.engfracmech.2007.03.027.

Saouma, V., Perotti, L., 2006. Constitutive Model for Alkali-Aggregate Reactions. Mater. J. 103 (3), 194–202.

Schlangen, E., Copuroglu, O., 2010. Modeling of expansion and cracking due to ASR with a 3D lattice model. 7th Int. Conf. Fract. Mech. Concr. Struct., 1017–1023

Shakoorioskooie, M., Leemann, A., Griffa, M., Zboray, R., Lura, P., 2019. Characterization of alkali-silica reaction damage in concrete by X-ray tomography. In Int. Conf. Tomogr. Mater. Struct.

Shi, Z., Geng, G., Leemann, A., Lothenbach, B., 2019. Synthesis, characterization, and water uptake property of alkali-silica reaction products. Cem. Concr. Res. ISSN 00088846. doi: 10.1016/j.cemconres.2019.04.009.

Stark, D., 1991. The Moisture Condition of Field Concrete Exhibiting Alkali-Silica Reactivity. Durab. Concr. Second Int. Conf. Montr. Canada 126, 973–988.

Suquet, P.M., 1985. Elements of Homogenization for Inelastic Solid Mechanics. In Homog. Tech. Compos. Media, chapter IV, pages 194–278.

Swamy, R.N., 1992. The Alkali-Silica Reaction in Concrete. Blackie and Son Ltd, Glasgow, London. https://doi.org/10.1016/0262-5075(86)90059-x. 0203036638.

Ulm, F.-J., Coussy, O., Kefei, L., Larive, C., 2000. Thermo-Chemo-Mechanics of ASR Expansion in Concrete Structures. J. Eng. Mech. 126 (March), 233–242.

Wriggers, P., Moftah, S.O., 2006. Mesoscale models for concrete: Homogenisation and damage behaviour. Finite Elem. Anal. Des. ISSN 0168874X. doi: 10.1016/j.finel.2005.11.008.

Wu, T., Temizer, I., Wriggers, P., 2014. Multiscale hydro-thermo-chemo-mechanical coupling: Application to alkali-silica reaction. Comput. Mater. Sci., 84: 381–395. ISSN 09270256. doi: 10.1016/j.commatsci.2013.12.029.

Xu, S., Zhu, Y., 2009. Experimental determination of fracture parameters for crack propagation in hardening cement paste and mortar. Int. J. Fract., 157 (1–2): 33–43. ISSN 03769429. doi: 10.1007/s10704-009-9315-x.